\documentclass[usenatbib,fleqn]{aa}  
\usepackage{amsmath}
\usepackage{color}
\usepackage{multirow}  
\usepackage[pdftex, linktocpage=true, colorlinks,citecolor=blue]{hyperref}
\hypersetup{
	colorlinks= true, 
	urlcolor= blue, 
	linkcolor=red, 
	citecolor=blue 
}
\usepackage{caption}
\usepackage{subcaption}
\newcommand\Tstrut{\rule{0pt}{2ex}}         
\newcommand\Bstrut{\rule[-0.5ex]{0pt}{0pt}} 

\usepackage{graphicx}
\usepackage{txfonts}
\begin{document} 
	
\title{Multi-wavelength campaign on NGC 7469}
\subtitle{VI. Photoionisation modelling of the emission line regions \\ and the warm absorber}

\author{S. Grafton-Waters \inst{\ref{MSSL}}\thanks{\email{sam.waters.17@ucl.ac.uk}} \and
	G. Branduardi-Raymont \inst{\ref{MSSL}} \and
	M. Mehdipour \inst{\ref{SRON}} \and
	M. J. Page \inst{\ref{MSSL}} \and
	E. Behar \inst{\ref{Haifa}} \and
	J. Kaastra \inst{\ref{SRON},\ref{Leiden}} \and
	N. Arav \inst{\ref{Virginia}} \and
	S. Bianchi \inst{\ref{Roma}} \and
	E. Costantini \inst{\ref{SRON}} \and
	J. Ebrero \inst{\ref{Madrid}} \and
	L. Di Gesu \inst{\ref{ASI}} \and
	S. Kaspi \inst{\ref{TelAviv}} \and
	G. A. Kriss \inst{\ref{Baltimore}} \and
	B. De Marco \inst{\ref{Warsaw}} \and
	J. Mao \inst{\ref{Glasgow},\ref{SRON}} \and
	R. Middei \inst{\ref{Roma}} \and
	U. Peretz \inst{\ref{Haifa}} \and
	P.-O. Petrucci \inst{\ref{Grenoble}, \ref{CNRS}} \and
	G. Ponti \inst{\ref{Garching}}
	}

\institute{
Mullard Space Science Laboratory, University College London, Holmbury St. Mary, Dorking, Surrey, RH5 6NT, UK \label{MSSL} \and 
SRON Netherlands Institute for Space Research, Sorbonnelaan 2, 3584 CA Utrecht, The Netherlands \label{SRON}  \and
Department of Physics, Technion-Israel Institute of Technology, 32000 Haifa, Israel  \label{Haifa} \and
Leiden Observatory, Leiden University, PO Box 9513, 2300 RA Leiden, The Netherlands \label{Leiden} \and 
Department of Physics, Virginia Tech, Blacksburg, VA 24061, USA \label{Virginia} \and
Dipartimento di Matematica e Fisica, Universit\`{a} degli Studi Roma Tre, Via della Vasca Navale 84, 00146 Roma, Italy \label{Roma} \and
European Space Astronomy Centre, PO Box 78, 28691 Villanueva de la Caada, Madrid, Spain \label{Madrid} \and
Italian Space Agency (ASI), Via del Politecnico snc, 00133, Roma, Italy \label{ASI} \and
School of Physics and Astronomy and Wise Observatory, Tel Aviv University, Tel Aviv 69978, Israel \label{TelAviv} \and
Space Telescope Science Institute, 3700 San Martin Drive, Baltimore, MD 21218, USA \label{Baltimore} \and
Nicolaus Copernicus Astronomical Center, Polish Academy of Sciences, Bartycka 18, PL-00-716 Warsaw, Poland \label{Warsaw} \and
Department of Physics, University of Strathclyde, Glasgow, G4 0NG, UK \label{Glasgow} \and
Univ. Grenoble Alpes, IPAG, F-38000 Grenoble, France \label{Grenoble} \and
CNRS, IPAG, F-38000 Grenoble, France \label{CNRS} \and
Max-Planck-Institut für Extraterrestrische Physik, Giessenbachstrasse, 85748, Garching, Germany \label{Garching} 
}

\date{Received 01 May 2019 / Accepted 03 July 2019}


\abstract
{}
{We aim to investigate and characterise the photoionised X-ray emission line regions within the Seyfert 1 galaxy NGC 7469.}
{We applied the photoionisation model, \texttt{PION}, within the spectral fitting code \texttt{SPEX} to analyse the 640 ks reflection grating spectrometer (RGS) spectrum of NGC 7469 gathered during an XMM-Newton observing campaign in 2015.}
{We find the emission line region in NGC 7469 to be multiphased, consisting of two narrow components with ionisation parameters of log $\xi =$ 0.4 and 1.6. A third, broad emission component, with a broadening velocity of $v_b \sim 1400$ km s\textsuperscript{-1} and an outflow velocity of $v_{out} \sim -4500$ km s\textsuperscript{-1} is required to fit the residuals in the \ion{O}{VII} triplet at around 22 \AA. Assuming a volume filling factor of 0.1, the lower distance limits of the narrow emission line region components are estimated for the first time at 2.6 and 2.5 pc from the central black hole, whereas the broad component has an estimated lower bound distance between 0.004 to 0.03 pc, depending on the assumed plasma parameters. The collisionally ionised plasma from the star burst region in NGC 7469 has a plasma temperature of 0.32 keV and an outflow velocity of -280 km s\textsuperscript{-1}, which is consistent with previous results in this campaign. 
In addition, we model the photoionised plasma of the warm absorber (WA) in NGC 7469 and find that it consists of three photoionised 
phases with different values of $\xi$, $N_H$ and $v_{out}$. 
The upper bound distances of these WA components are 1.9, 0.3, and 0.6 pc, respectively, consistent with archival results.} 
{The environment of NGC 7469 is a complex mix of plasma winds absorbing and emitting X-rays. We find the picture painted by our results can be attributed to line emitting plasma located at distances ranging from near the black hole to the torus and beyond the ionised outflows.}

\keywords{X-rays: 
	Galaxies --
	Galaxies: 
	Active --
	Galaxies: 
	Seyfert --
	Galaxies: 
	Individual: 
	NGC 7469 --
	Technique: 
	Spectroscopic 
}

\titlerunning{Photoionisation Modelling of the Emission Line Regions in NGC 7469}
\authorrunning{Grafton-Waters et al.}
\maketitle


\section{Introduction}

Active galactic nuclei (AGN) are one of the most extreme environments in the universe, powered by the accretion of matter into the central super massive black hole (SMBH). Matter is also ejected away from the AGN, into the interstellar medium (ISM) of the host galaxy. The coupling of the outflowing matter from the SMBH with the surrounding galaxy is known as feedback, and is thought to be responsible for the coevolution of the black hole and its host galaxy \citep[e.g.][]{DiMatteo2005,Hopkins2010, Soker2011}. The black hole mass correlates exceptionally well with the velocity dispersion of the stars in the galaxy bulge \citep{Ferrarese2000, Gebhardt2000}.

The outflowing ionised wind within an AGN is known in the X-ray literature, and similarly in the UV band, as the warm absorber (WA). Although the origins and launching mechanisms are not fully understood, there is evidence to suggest that the WA wind is part of a large scale outflow, with an ultra fast wind found closer to the black hole, originating from the accretion disc \citep{Risaliti2005, Cappi2006, Tombesi2010, Tombesi2012}, and the less dynamic WA wind stemming from the torus \citep{Blustin2005}.


Warm absorbers were first observed by \cite{Halpern1984}, and have since been found to exhibit strong, narrow absorption lines \citep[e.g.][]{Kaastra2000, Kaastra2002}, with blueshifted velocities between about 100 - 1000 km s\textsuperscript{-1}. Around half of Seyfert galaxies contain WA winds \citep[e.g.][]{Reynolds1995} which are multi-temperature \citep[e.g.][]{Krolik2001} and multi-phased \citep[e.g.][]{Kaastra2002, Kaastra2014, Behar2017, Mehdipour2018}.

Warm absorbers are ionised through photoionisation processes, but modelling photoionised plasma is very complex due to the various interacting processes within the plasma winds \citep[see e.g.][]{Mehdipour2016}. However, the ionisation and thermal states of the photoionised plasma can be quantified as a single parameter, the ionisation parameter $\xi$. The ionisation parameter is given by
\begin{equation}
\xi = \frac{L_{ion}}{n r^2},
\label{XI_EQ}
\end{equation}
where $L_{ion}$ is the ionising luminosity between 13.6 eV and 13.6 keV (1 - 1000 Ryd), \textit{n} is the hydrogen number density and \textit{r} is the distance of the photoionised gas from the central X-ray source. The ionisation parameter (through photoionisation modelling) and the luminosity (via spectral energy distribution modelling), can both be measured. However, n and $r^2$ are degenerate with respect to each other, preventing a direct derivation of the WA distance. Methods to find \textit{n} and \textit{r} depend on the recombination time (function of \textit{n}), which measures how fast the plasma in the WA responds to changes in the X-ray ionising luminosity \citep[e.g][]{Kaastra2012, Ebrero2016}. They also heavily rely on limiting the density and the location from the long-term changes in the WA \citep{Mehdipour2018}. Consequently, these methods require variability in the X-ray source. Alternatively, constraints on the density or distance of the WA can be derived via density sensitive metastable absorption lines \citep{Mao2017}, which do not require X-ray variability in the source.

The broad and narrow line regions (BLR, NLR) give rise to the emission features found in AGN spectra. The optical spectrum of the BLR clouds (with densities $n_{BLR} > 10^{15}$ m\textsuperscript{-3}) show broad emission lines with Doppler velocity broadening of a few thousand km s\textsuperscript{-1} \citep[e.g.][]{Peterson1997, Krolik1999, Netzer2015}. The BLR is likely to be located at the outer part of the accretion disk, where the temperature is low enough for dust to form; however, this dust is broken apart into cloudlets by the radiation from the central source \citep[see e.g.][]{Czerny2011, Czerny2015, Czerny2016, Czerny2017}. 
On the other hand, a continuous wind could be achieved if the BLR is shielded from the radiation by some dense gas closer to the central source \citep{Murray1995}. However, for an opposing viewpoint, see \cite{Hamann2013}. In comparison, the NLR is located further out than the torus and has velocity broadening of a few hundred km s\textsuperscript{-1} and number density much less than that of the BLR \citep[$n_{NLR} \sim 10^{12}$ m\textsuperscript{-3}; e.g.][]{Netzer1990, Xu2007}. 

The NLR region is a multiphased plasma with different ionisations, extending over vast distances such that both soft X-ray and optical [\ion{O}{III}] emissions can come from the same emission line region \citep{Bianchi2006}. The strongest emission features in the soft X-ray band are those from the He-like \ion{O}{VII} triplet \citep[e.g][]{Whewell2015, Mao5548, Behar2017}. In addition, the Ly$\alpha$ lines of \ion{C}{VI} and \ion{O}{VIII}, and subsequent species, are prominent in many X-ray spectra, as are radiative recombination continuum (RRC) features \citep[e.g.][]{Blustin2007, Whewell2015, Behar2017}. 

The distances to the optical and X-ray NLRs in the heavily studied AGN NGC 5548 have been calculated between 1 to 3 pc \citep{Peterson2013} and 1 to 15 pc \citep{Detmers2009}, respectively. From photoionisation modelling, \cite{Whewell2015} derived properties of the NLR ($\log \xi = 1.5$ and $\log N_H = 22.9$ m\textsuperscript{-2}) and computed its distance at 13.9 pc. As a comparison, the WA in NGC 5548 is located between 5 to 10 pc \citep{Ebrero2016}. Interestingly, the X-ray emission line region components in NGC 5548 have been found to have blueshifted velocities between -50 and -400 km s\textsuperscript{-1} \citep[a range of ionisation parameters were found at $\log \xi \sim 0.1 - 1.3$;][]{Mao5548}. In NGC 7469, the two X-ray emission line region components, with $\log \xi =$ 1 and -1, have outflow velocities of $v_{out} \simeq -470$ km s\textsuperscript{-1} \citep{Behar2017}. Similarly, in Mrk 509, the ion lines in the UV emission line region have a range of blueshifted and redshifted velocities from -300 to +600 km s\textsuperscript{-1} \citep{Kriss2019}.

NGC 7469 is a type I Seyfert AGN, at a redshift of $z = 0.016268$ \citep{Springob2005} and with a SMBH of mass $M_{BH} = 1 \times 10^{7} M_{\sun}$ \citep{Peterson2014}. NGC 7469 also contains a star burst ring, which circles the nucleus \citep[e.g][]{Wilson1991, Mehdipour2018}. Active galaxies with both photoionised plasma and a star burst region are not uncommon. For example, NGC 1365 has been found to contain two collisionally ionised dominant plasma regions and three, weaker, photoionised components \citep{Guainazzi2009, Whewell2016}. 

Previous X-ray investigations on NGC 7469 have found the X-ray WA to contain three phases, with ionisation parameters log $\xi$ = 0.8, 2.7, and 3.6 (erg cm s\textsuperscript{-1}), and two velocity domains -580 to -720, and -2300 km s\textsuperscript{-1} \citep{Blustin2007}. In the UV regime, two absorption components were found in NGC 7469, with kinematic properties of -570 and -1900 km s\textsuperscript{-1} \citep{Kriss2003, Scott2005}, with the former component having a velocity comparable to the high ionisation X-ray component found with Chandra \citep{Scott2005}. 

Recently, a multiwavelength campaign of NGC 7469, using XMM-Newton, Chandra, HST, Swift, and NuSTAR (along with ground based telescopes), was undertaken in 2015 \citep{Behar2017}. The 640 ks reflection grating spectrometer (RGS) spectrum was analysed by \cite{Behar2017} and \citet{Peretz2018} and the 237 ks high-energy transmission grating spectrometer (HETGS) spectrum from Chandra was investigated by \cite{Mehdipour2018}. From studying the column density variability over ten years and finding no change, \citet{Peretz2018} constrained the lower distance limit of the WA ($r > 12 - 31$ pc), concluding that the outflow wind was located far from the X-ray source. This is consistent with Chandra data where the upper limit of the WA was found at $r < 80$ pc \citep{Mehdipour2018}. In the same campaign, three UV components have been found at -530, -1420, and -1900 km s\textsuperscript{-1}, where the distances of components 1 and 3 were found at 6 and between 60 - 150 pc, respectively \citep{Aravprep}. The broadband X-ray spectrum of NGC 7469 has been explained in terms of a two coronal model \citep[see e.g.][]{Petrucci2013}: the primary emission was consistent with Comptonisation of photons in a  hot, optically thin electron corona ($T_{e,\ hot} = 45^{+15}_{-12}$ keV) while a warm, optically thick corona ($T_{e,\ warm} = 0.67 \pm 0.03$ keV) was found to best fit the soft excess \citep{Middei2018}.  

In this paper, we model the emission lines in the same 640 ks RGS spectrum of NGC 7469 analysed by \cite{Behar2017}, obtained during the multiwavelength campaign in 2015. The stacked RGS spectrum consists of seven observations, each an average of 90 ks, from both RGS 1 and RGS 2. \citep[See][for the observation log, Table 1, and details of data reduction via the XMM-Newton SAS pipeline.]{Behar2017} The emission line features have been modelled previously in NGC 7469 \citep{Blustin2007, Behar2017}, but here we carry out a physically self-consistent analysis using the recently developed code \texttt{PION} in \texttt{SPEX}. \texttt{PION} is the only model currently available that computes explicitly all the absorption and emission lines in the spectrum, computing the photoionisation balance on the fly using the continuum fit as the ionising spectral energy distribution, allowing for the determination of $\xi$ and $N_H$ from a simple fitting session. Before the implementation of \texttt{PION}, photoionisation modelling had to be manually computed using a grid of parameters \citep[e.g.][]{DiGesu2013, Whewell2015}. The emission line regions in Seyfert 1 AGN have been investigated in great detail in NGC 5548 \citep{Whewell2015, Mao5548} and more recently in NGC 3783 \citep{Mao3783}. In this paper, we focus on the emission line regions of NGC 7469 and measure, for the first time, their location with respect to the central black hole. In addition, we also reanalyse the absorption features in the RGS spectrum, produced by the WA wind. Our approach makes physically self-consistent assumptions in the photoionisation modelling, giving a more physical insight; this represents an important improvement from the previous phenomenological fit of the RGS spectrum \citep{Behar2017}. 

In this paper, Section \ref{Data_Analysis} outlines the analysis and spectral modelling of the RGS spectrum, with the results shown in Section \ref{Results}. The examination of the \ion{O}{VII} triplet and the locations of both WA and emission line components are discussed in Section \ref{EM_Characteristics}. We discuss the AGN structure in Section \ref{Discussion} and present our conclusions in Section \ref{Conclusion}.

\section{Data analysis \label{Data_Analysis}}

We modelled the photoionised plasma in NGC 7469, fitting for both absorption and emission features, using \texttt{SPEX} \citep[\texttt{v3.04.00};][]{Kaastra1996}. We analysed the RGS data between 7 and 38 \AA\ and binned by a factor of 2. For statistical fitting, we used the Cash statistic \citep[C-statistic hereafter;][]{Cash1979}, with errors given at 1$\sigma$ confidence level. Finally, solar-abundances of \cite{Lodders2009} were used, as was a redshift of $z = 0.016268$ \citep{Springob2005} for NGC 7469.
\subsection{Spectral energy distribution}
\label{Sec_SED}
The broad band ionising continuum (optical-UV-X-ray), which we have adopted to fit the RGS spectrum, was modelled using the spectral energy distribution (SED) derived in \cite{Mehdipour2018}. The hard X-ray continuum was modelled by a power-law (\texttt{POW} in \texttt{SPEX}), produced when the photons from the disc are Compton up-scattered in the hot electron corona. Phenomenologically, the high energy emission of the source had been found to be consistent with a high energy cut-off of $E_{cut} = 170^{+60}_{-40}$ keV \citep{Middei2018}, which we applied to the power law. A warm optically thick medium was found to best fit the soft excess and explain the UV and optical photons from the disc \citep{Middei2018}, so we modelled this with a warm Comptonisation component \citep[\texttt{COMT};][]{Titarchuk1994}, as described by \cite{Mehdipour2015, Mehdipour2018}. Finally, the \ion{Fe}{K}$\alpha$ line and Compton hump were modelled with a reflection component \citep[\texttt{REFL};][]{Magdziarz1995}. Although the \texttt{REFL} and high energy cutoff are outside the RGS energy range, \texttt{PION} uses the full SED continuum model to calculate both the absorption and emission spectrum and ionisation and thermal balance of the plasma at the same time, self-consistently. Therefore, including the  full broad band ionising continuum allows \texttt{PION} to achieve more accurate photoionisation and ionisation balance results \citep{Mehdipour2016}.

 The parameters for these three SED components were initially set to the values found by \cite{Mehdipour2018}, 
with the normalisation ($N_{pow}$) and photon index ($\Gamma$) of \texttt{POW}, and the normalisation ($N_{comt}$) and electron temperature ($T_{e}$) of \texttt{COMT} let free to fit to the RGS spectrum; the best fit values are shown in Table \ref{SED_Results}. The adopted SED fitted to the RGS spectrum agrees with the EPIC-PN spectrum within 10\% in the RGS range (0.35 - 1.8 keV), 30\% between 2 and 6 keV and 50\% between 6 and 10 keV. The initial C-statistic, without any WA and soft X-ray emission line models, was C = 5212 (for 1546 degrees of freedom; d.o.f hereafter). 

\subsection{Galaxy absorption}
\label{Galactic_Absorption}

The absorption through the Galaxy was taken into account throughout the spectral fitting using the \texttt{HOT} component in \texttt{SPEX}. Initially we set the Galactic column density to $N^{Gal}_{H} = 5.5 \times 10^{24}$ m\textsuperscript{-2} \citep{Wakker2006, Wakker2011}, which we allowed to fit at the same time as the WA components (see Table \ref{SED_Results}). The temperature for neutral gas was fixed at $T_{Gal} = 0.5$ eV and the turbulent velocity was fixed at $v^{Gal}_{turb} = 5.64 \pm 0.96$ km s\textsuperscript{-1} (see Appendix \ref{Appendix} for details of obtaining this velocity value; no significant $\Delta C$ is achieved for a zero-turbulent velocity). The C-statistic, when fitting the SED with the addition of Galaxy absorption, was 3631 (for 1545 d.o.f).

In addition to this neutral gas in the Galaxy, \cite{Behar2017} also identified ionised absorption from warmer gas in the ISM with a temperature of about 130 eV and column density equal to $2 \times 10^{23}$ m\textsuperscript{-2}. Therefore, we added a second \texttt{HOT} component with these parameter values, which were fixed throughout the spectral modelling (there is no significant $\Delta C$ when these parameters were fitted).

\subsection{Spectral modelling}
The RGS spectrum is rich in narrow absorption and emission lines. After fitting the underlying optical-UV-X-ray continuum (SED) and Galaxy absorption, the spectral features were modelled. Although the main aims of this paper are to analyse the emission line regions, we first of all have to model the absorption features in the RGS spectrum. Here, the specialist photoionisation model \texttt{PION} was used to analyse the photoionised plasma of both the WA and the emission line regions in NGC 7469. For absorption, the absorption covering factor (\texttt{$f_{cov}$}) was set at unity. For emission, the emission covering factor ($C_{cov} = \Omega/4\pi$) was allowed to vary between 0 and 1 (the absorption covering factor was set to $f_{cov} = 0$ when fitting the emission). 

\subsubsection{Absorption \label{Abs_Data}}
The WA components were applied one at a time to the spectrum, with initial values from Table 2 of \cite{Behar2017}, in order of decreasing contribution to $\Delta C$. We found that three absorption components fit the absorption features in the RGS spectrum, with each fit giving a significant change in the C-statistic value (see Table \ref{Absorption_Results}).  The C-statistic after all three WA components were initially fitted is C = 2780 (for 1538 d.o.f). A fourth component only improved the statistical best fit by $\Delta C = 12$, significantly less than the other three components (see Table \ref{Absorption_Results}). After we fit the data with all four WA components, we found that this fourth component had the same ionisation parameter as component 3, but with a different blueshifted velocity of $v_{out,\ 4} \sim -920$ km s\textsuperscript{-1}, compared to $v_{out,\ 3} \sim -2100$ km s\textsuperscript{-1}. Therefore, as this component produced the same absorption lines as component three, except with a different line strength and blueshifted position, and since the C-statistic has changed little introducing it, we did not include a fourth WA component in the spectral fitting. Further to this, we tried to add WA components 1 and 6 from \cite{Behar2017} to the fit using \texttt{PION}. However, due to the very small column densities of these two components, no absorption features were detected. Therefore, we conclude that we require three WA components to describe the absorption features in the RGS spectrum, instead of the six found by \cite{Behar2017}. The model discrepancies between these two papers are not beyond what is expected from different analysis codes \citep{Mehdipour2016}; \citet{Behar2017} used the photoionised plasma code \texttt{XSTAR} in their \texttt{XSPEC} fits. 

The ionisation parameters of all three WA components were constrained to fit between $0 \leq \log \xi \leq 3$, to reduce the possibility that \texttt{SPEX} tries to find a best fit which cannot be constrained by the RGS operational range. From fitting these components, we found that WA component 2 requires $\log \xi = 3$, which is the upper limit of the set range. Therefore, we fix the ionisation parameter value at $\log \xi = 3$ for component 2, which also reduced the number of free parameters. 

Before fitting all the emission components, we freed all the parameters of interest (the three WA components, the column density for neutral Galactic absorption and the four SED parameters - $N_{pow}$, $\Gamma$, $N_{comt}$ and $T_e$) and fitted them together. This improved the C-statistic further to a new best fit value of C = 2737 ($\Delta C = 43$) for 1533 d.o.f.

\subsubsection{Emission}
The emission features in the RGS spectrum were fitted in much the same way as the WA components, with the addition of the emission covering factor ($C_{cov}$) being a free parameter. Two narrow emission components (EM1 and EM2) were fitted one at a time, improving the C-statistic by $\Delta C_{EM1} = 230$ and $\Delta C_{EM2} = 34$, respectively (C = 2473 for 1541 d.o.f). A third narrow component did not have an effect on the statistical best fit. 

After fitting the two narrow emission components, we detected an excess fit residual in the \ion{O}{VII} triplet at 22 \AA, which required an additional component. A broad emission component was found to significantly improve the fit to the spectra of NGC 5548 \citep{Mao5548} and  NGC 3783 \citep{Mao3783}, therefore, a third \texttt{PION} emission component (EM3), coupled with a Gaussian velocity broadening component \texttt{VGAU} (represented by the parameter $v_b$), was also fitted. The broad component improves the fit by $\Delta C =23$, reducing the residuals at 22 \AA\ (C = 2450 for 1545 d.o.f). 

\begin{table}
	\centering
	\caption{Final best fit parameter values for the \texttt{POW}, \texttt{COMT}, and \texttt{HOT} components (see Sections \ref{Sec_SED} and \ref{Galactic_Absorption} for details).}
	\label{SED_Results}
	\begin{tabular}{c | c c}
		\hline
		\hline
		Component & Parameter & Value \\
		\hline
		\texttt{POW}& $N_{pow}$\ \ \tablefootmark{a} & $6.70 \pm 0.01$ \Tstrut\Bstrut \\
		& $\Gamma$ & $2.07 \pm 0.01$ \\
		\hline
		\texttt{COMT} & $N_{comt}$\ \ \tablefootmark{b} & $8.81^{+0.12}_{-0.35}$ \Tstrut\Bstrut \\
		& $T_e$ (keV)& $0.118 \pm 0.001$\\
		\hline
		\texttt{HOT} & $N_{H}^{Gal}$\ \ \tablefootmark{c}& $3.66^{+0.03}_{-0.02}$ \Tstrut\Bstrut \\ 
		& $v^{Gal}_{turb}$ (km s\textsuperscript{-1})\ \ \tablefootmark{d} & $5.64 \pm 0.96$ \Tstrut\Bstrut \\ 
		\hline
	\end{tabular}\\
	\tablefoot{
		\tablefoottext{a}{$\times 10^{51}$ photons s\textsuperscript{-1} keV\textsuperscript{-1} at 1 keV;}
		\tablefoottext{b}{$\times 10^{57}$ photons s\textsuperscript{-1} keV\textsuperscript{-1};}
		\tablefoottext{c}{$\times 10^{24}$ m\textsuperscript{-2};}
		\tablefoottext{d}{fixed, see Appendix \ref{Appendix} for details.}}
\end{table}

\subsubsection{Nuclear star burst ring}
\label{Sec_CIE}
Further to the photoionised emission, emission from collisionally ionised equilibrium (CIE) plasma, within the star burst region, was modelled using the \texttt{CIE} component in \texttt{SPEX}. The electron temperature ($T_{e}$), emission measure (EM) and outflow velocity ($v_{out}$) were the fitted parameters. The CIE component improved the global C-statistic by $\Delta C = 9$. This does not seem very significant, compared to the other components, but without this component, the star burst emission lines of \ion{Fe}{XVII} at 15.3 and 17.4 \AA\ \citep{Behar2017} would not be accounted for; hence, we include this component in our best fit nevertheless. 

\subsection{Abundances} 
\label{Sec_Abund}
After refitting all the \texttt{PION} emission and \texttt{CIE} components together (improving the statistical fit only by $\Delta C = 11$), we let the abundances of C, N, Ne, Mg, S and Fe free in WA component 1 and refitted. The abundances were calculated with respect to Oxygen as no Hydrogen lines are present in the RGS spectrum \citep{Behar2017}. The abundances of the remaining absorption and emission components (both \texttt{PION} and \texttt{CIE}) were coupled to WA component 1 and are displayed in Table \ref{Absorption_Results}. This is because we assumed that the chemical enrichment was the same throughout the AGN nucleus and star burst region. The abundances are consistent with the values, within errors, from \cite{Behar2017}, except the abundance of S is three times larger in this paper. The C-statistic decreased by $\Delta C = 54$ after freeing the above ion abundances (C = 2376 for 1544 d.o.f). 

\section{Spectral fitting results \label{Results}}

\begin{figure*}
	\centering
	\begin{subfigure}{0.95\linewidth}
		\includegraphics[width=1\linewidth]{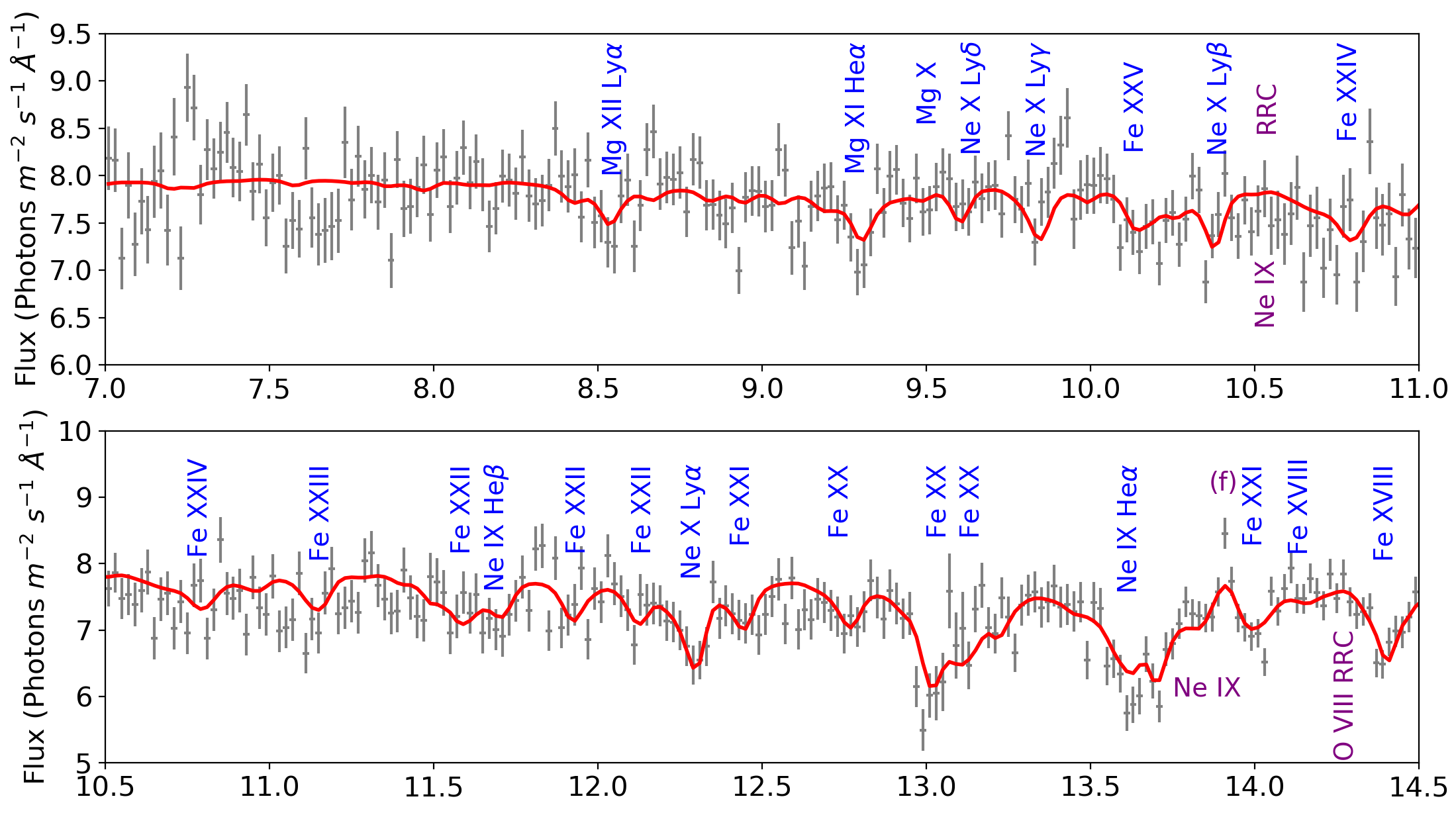}
	\end{subfigure}
	\begin{subfigure}{0.95\linewidth}
		\includegraphics[width=1\linewidth]{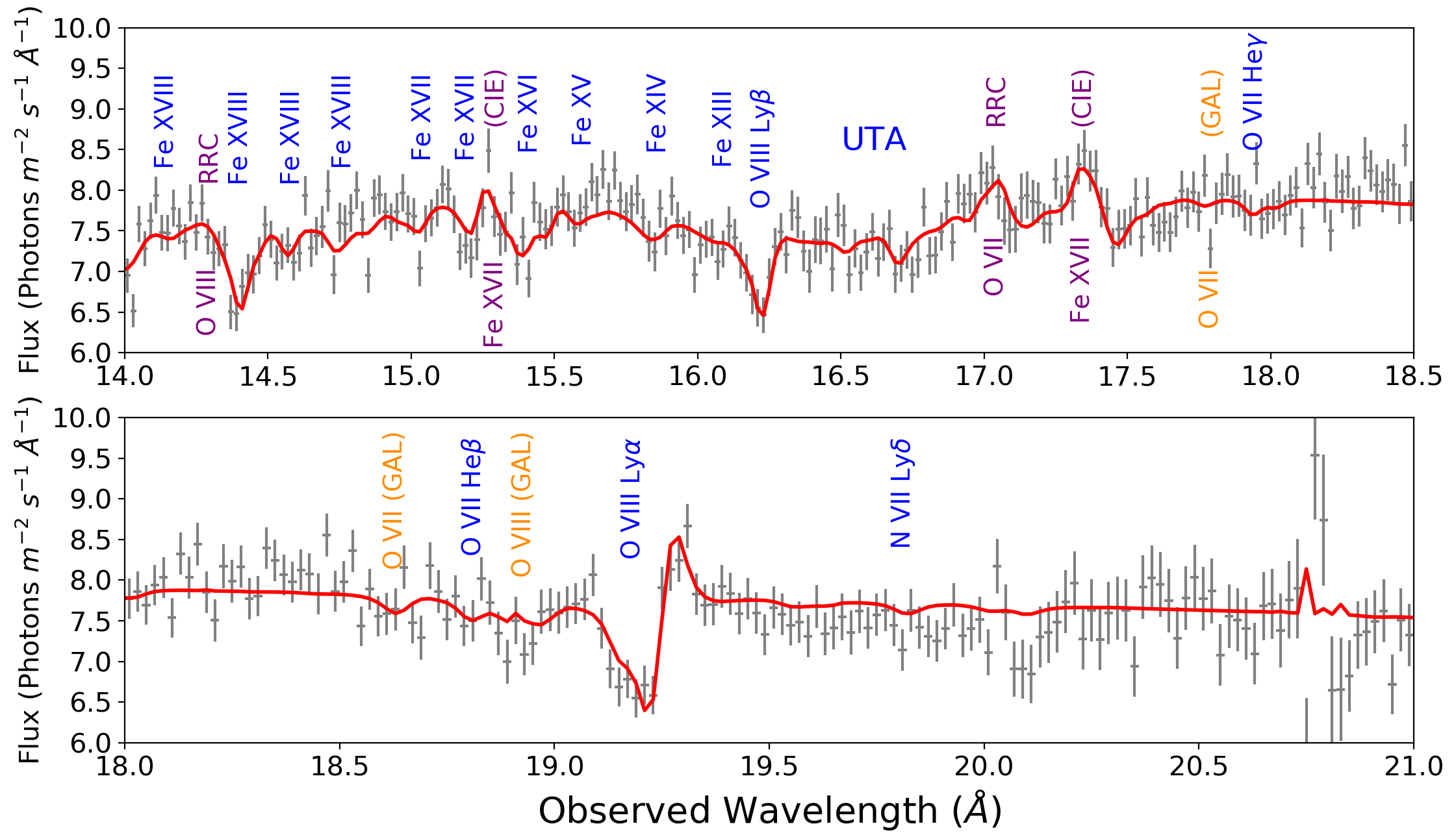}
	\end{subfigure}
	\caption{RGS spectrum of NGC 7469 in the observed frame. The best fit model (red line) is over plotted. The significant absorption and emission features are labelled with their corresponding ions (blue and purple, respectively). Galactic absorption features are indicated in orange.}
	\label{PION_SPEC_1}
\end{figure*}

\begin{figure*}
	\centering
	\begin{subfigure}{0.95\linewidth}
		\includegraphics[width=1\linewidth]{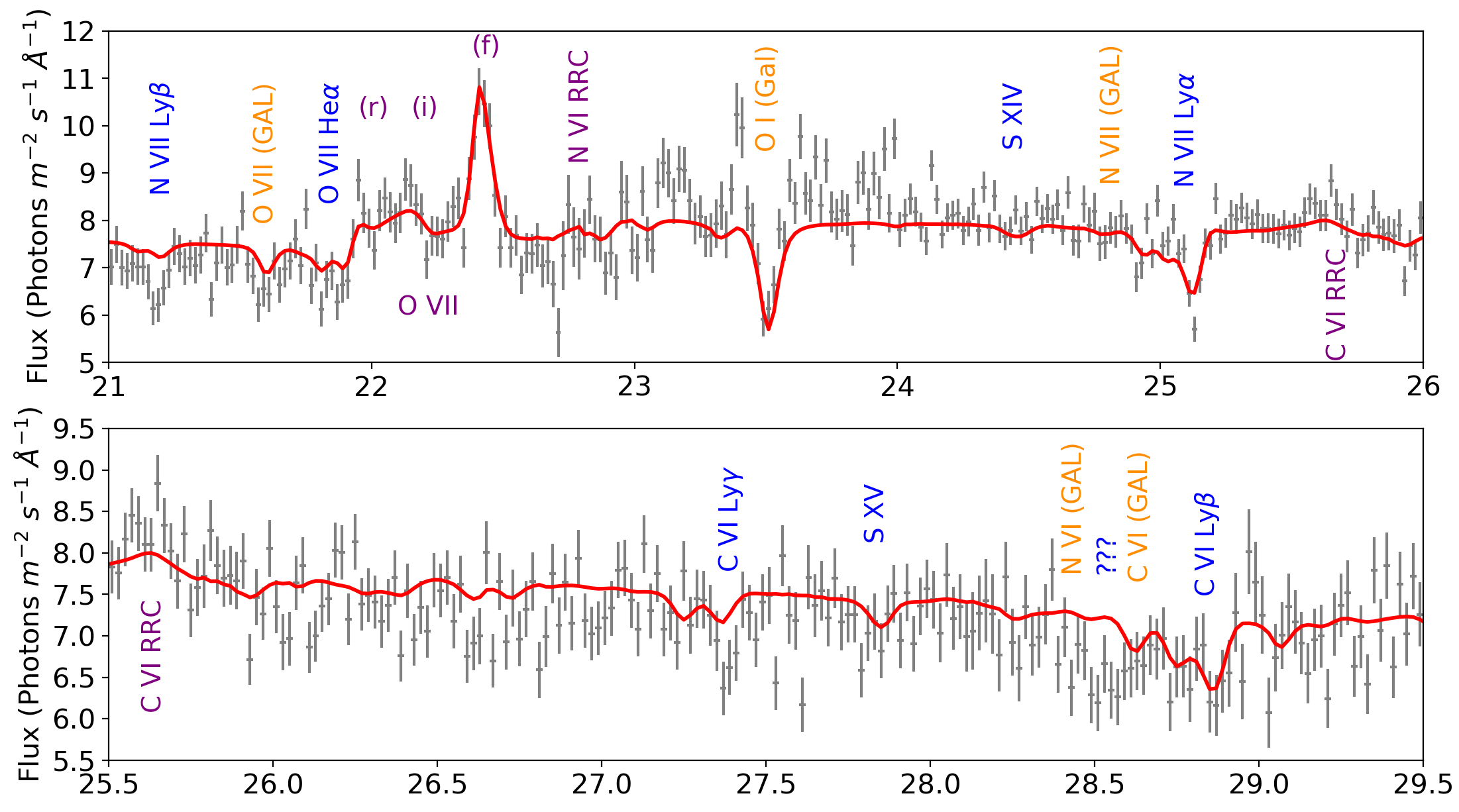}
	\end{subfigure}
	\begin{subfigure}{0.95\linewidth}
		\includegraphics[width=1\linewidth]{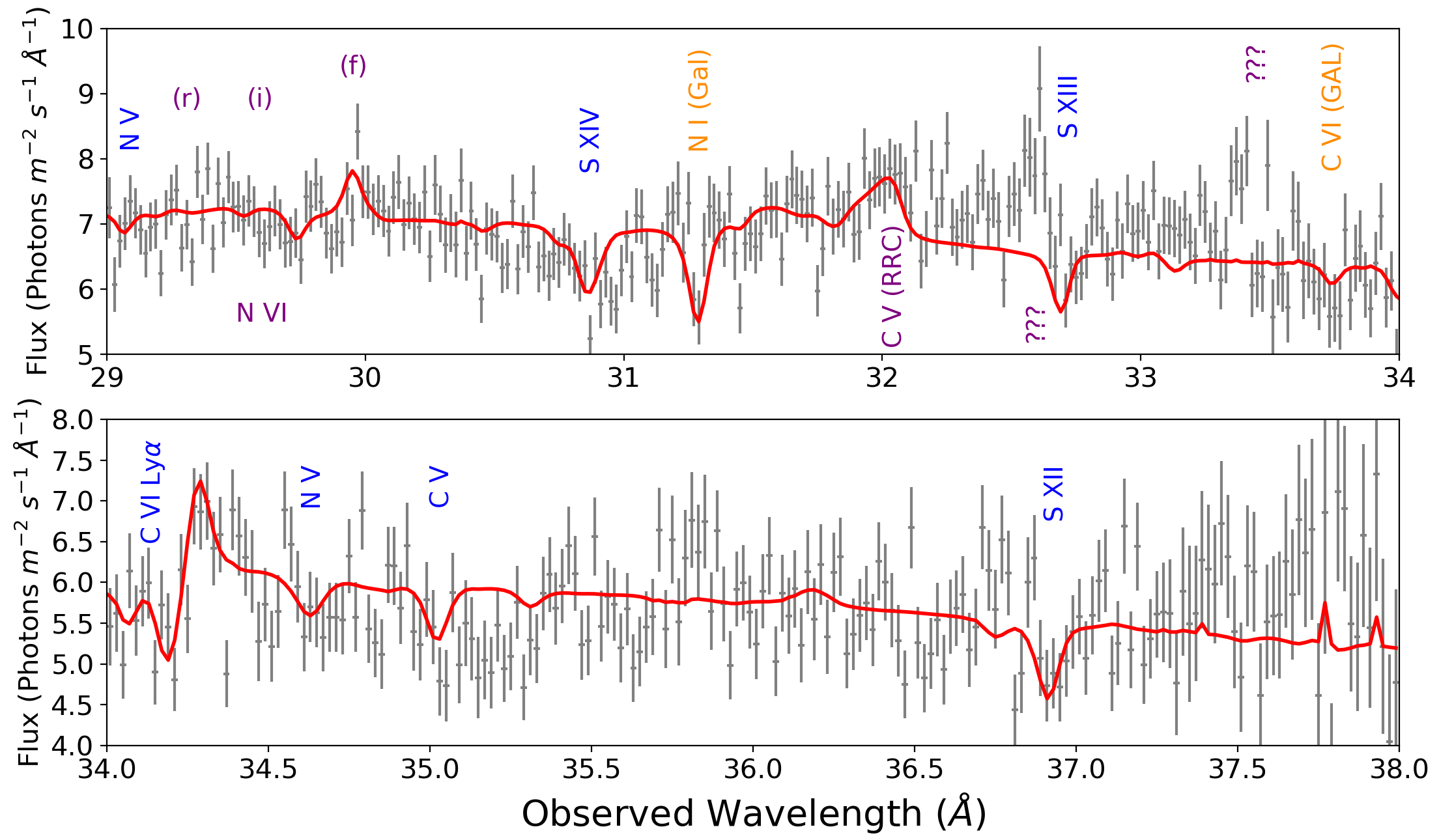}
	\end{subfigure}
	\caption{Figure \ref{PION_SPEC_1} Continued.}
	\label{PION_SPEC_2}
\end{figure*}

The RGS spectrum and best fit model are displayed in Figures \ref{PION_SPEC_1} and \ref{PION_SPEC_2}, with significant absorption and emission features labelled with their respective ions; all wavelengths are in the observed reference frame. Table \ref{SED_Results} displays the best fit parameter values from the \texttt{POW} and \texttt{COMT} components, in addition to the parameters of the neutral \texttt{HOT} component. 

The final model best fit C-statistic value is a fairly acceptable 2252 for 1512 d.o.f, obtained by fitting all parameters together. However, the expected C-statistic, calculated by \texttt{SPEX} \citep{Kaastra2017} is $1551 \pm 56$. The large difference could be due to the unaccounted for emission and absorption features in the spectrum (Figure \ref{PION_SPEC_2}, for example at 28.5, 32.6 and 33.4 \AA) by the current best fit. \cite{Behar2017} found no physical explanation for these emission lines in terms of ionic species, so we try to fit these features with a relativistically broadened emission line from the accretion disc \citep[][see Section \ref{Laor_Sec}]{Laor1991}. Another explanation for this high C-statistic value may be RGS calibration issues, whereby the effective area requires some time and wavelength dependent corrections, up to 10\% in some cases \citep{Kaastra_RGS}. On the other hand, a thorough understanding of the physical models outweighs the drive to obtain a further minimised statistical best fit \citep{Blustin2002}.

\subsection{The photoionised wind}
\begin{figure}
	\centering
	\begin{subfigure}{1\linewidth}
		\includegraphics[width=\linewidth]{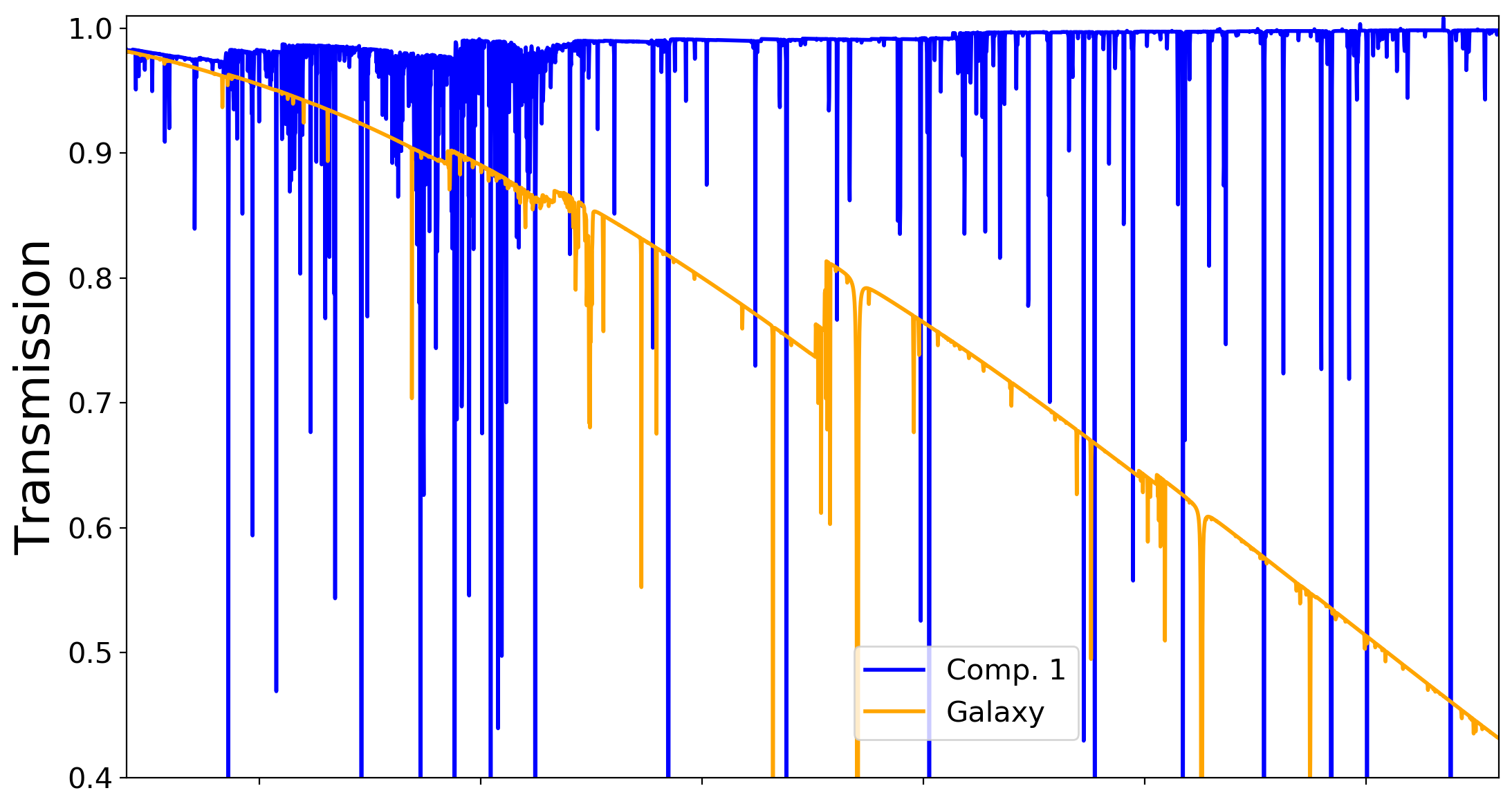}
	\end{subfigure}
	\begin{subfigure}{1\linewidth}
		\includegraphics[width=\linewidth]{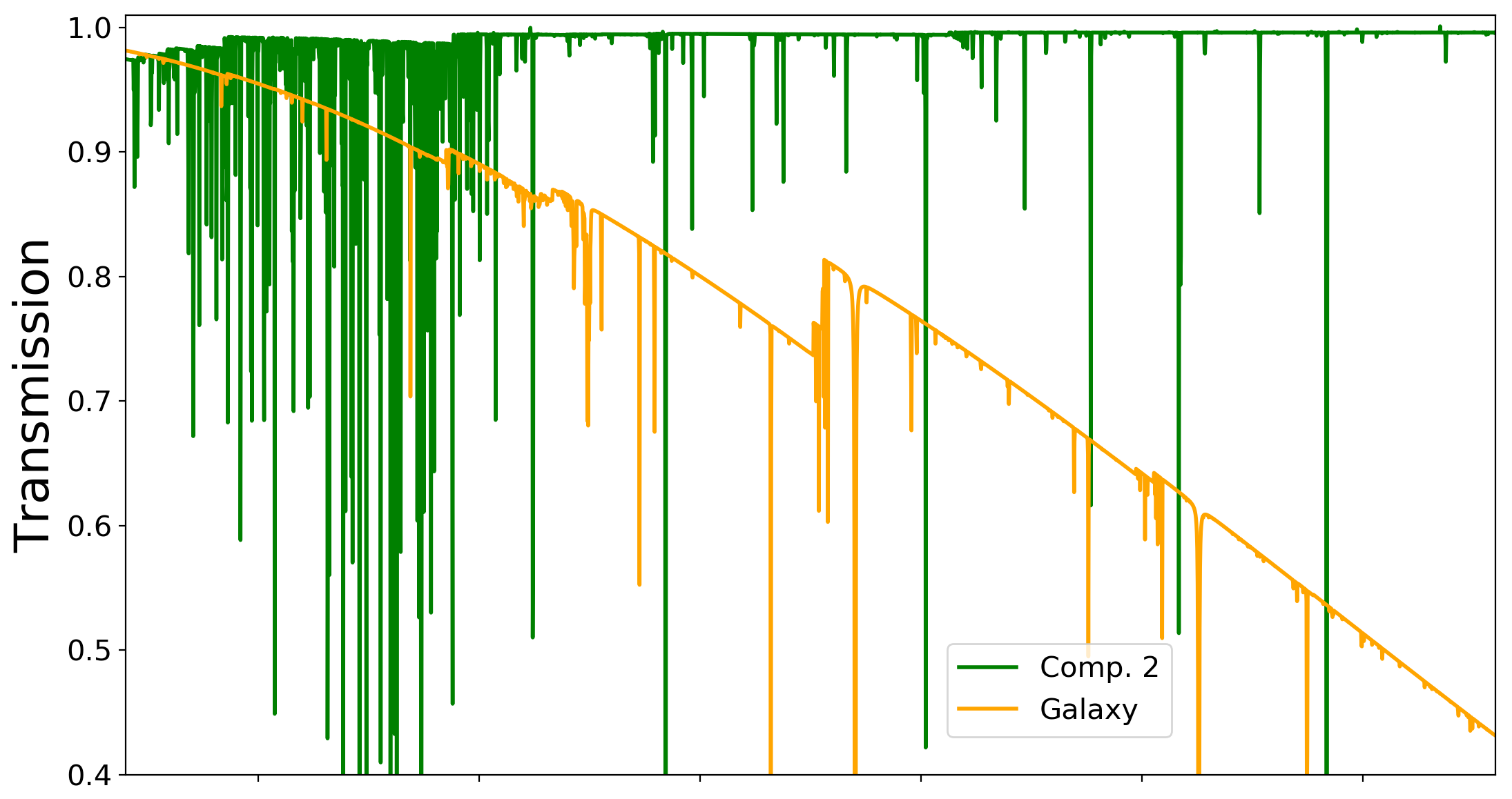}
	\end{subfigure}
	\begin{subfigure}{1\linewidth}
		\includegraphics[width=\linewidth]{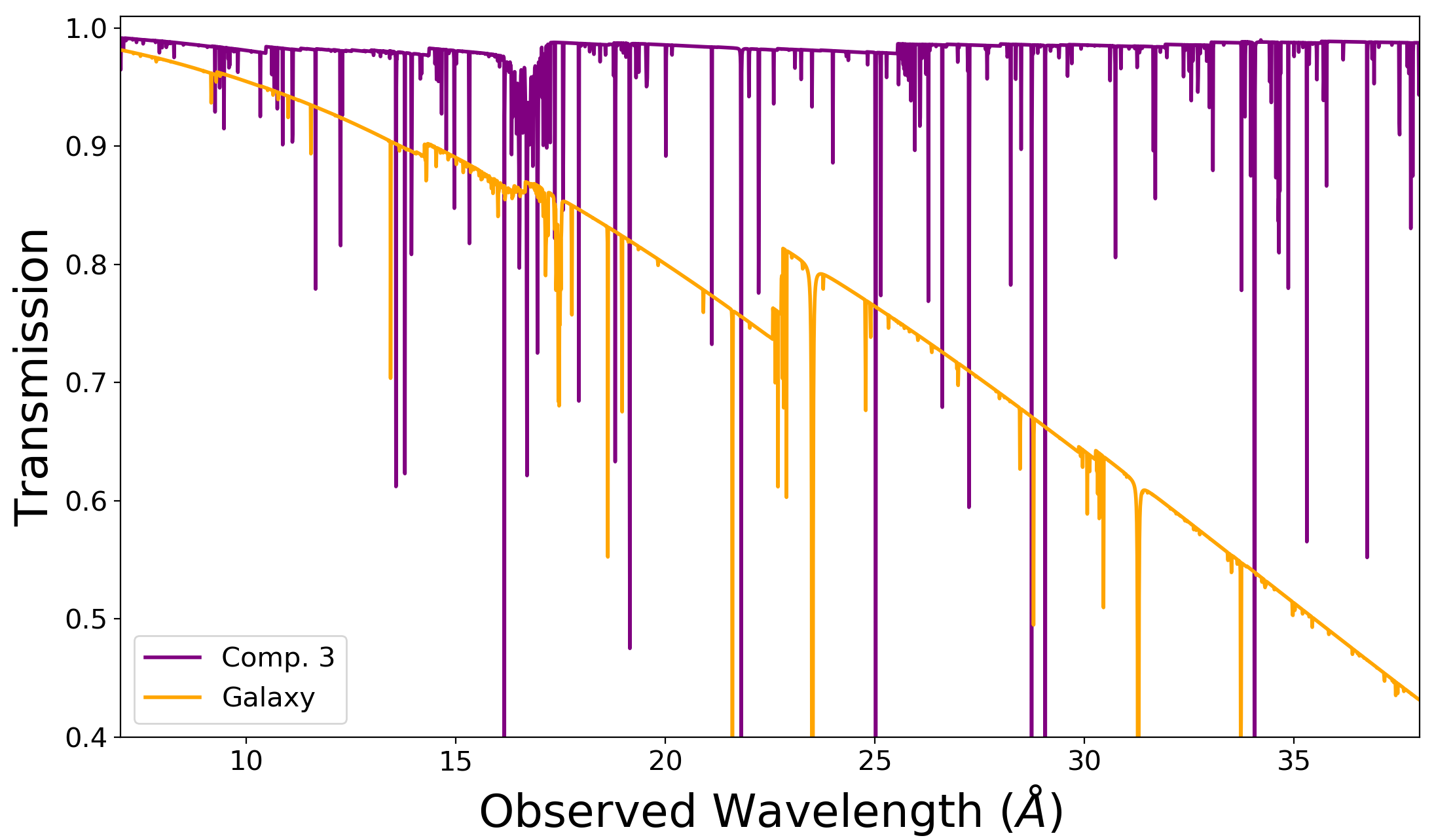}
	\end{subfigure}
	\caption{Plot of \texttt{PION} absorption models for each warm absorber component (1 - 3 from top to bottom). These plots help demonstrate what the model looks like with the parameter values from Table \ref{Absorption_Results}: large $\xi$ means absorption at higher energies; large $N_H$ means stronger absorption lines. The Galaxy absorption transmission is overlayed in orange for a comparison. The Galaxy absorption affects the continuum emission, especially at longer wavelengths, causing the transmission to decrease. As a consequence of the RGS resolution, weak narrow lines are not evident in Figure \ref{PION_SPEC_1} due to the low equivalent width, but can be seen in these models.}
	\label{Model_WAs}
\end{figure}

The best fit WA parameters and abundances of the six ions are displayed in Table \ref{Absorption_Results}. The WA components have a range of equivalent hydrogen column densities: the total column density of the WA is $N^{Tot}_{H} = 64 \times 10^{24}$ m\textsuperscript{-2}, approximately two times larger than the total equivalent hydrogen column density found by \cite{Behar2017}. These WA components have three ionisation phases ranging from log $\xi =$ 1.5 - 3.0 and three kinematic phases, consistent with the X-ray kinematics from Table 2 in \cite{Behar2017} and UV velocities \citep{Kriss2003, Scott2005, Aravprep}. Figure \ref{Model_WAs} illustrates the absorption lines produced by each of the three WA model components, in the observed reference frame. The Galactic absorption model (orange line) shows how much of the continuum is absorbed, especially at longer wavelengths.

Component 1, log $\xi = 2.32 \pm 0.01$, is responsible for the H-like and He-like ions of \ion{Ne}{IX}, \ion{Ne}{X} and \ion{O}{VII}, as well as the absorption lines of \ion{C}{V} - \ion{C}{VI}, \ion{S}{XII} - \ion{S}{XV} and \ion{Fe}{XIII} - \ion{Fe}{XVIII}. Component 2, log $\xi$ = 3.00, accounts for the H-like and He-like ions of \ion{Mg}{XI}, \ion{Mg}{XII}, \ion{Ne}{IX} and \ion{Ne}{X}, along with the \ion{Fe}{XV} - \ion{Fe}{XVIII} absorption lines. Component 3 is the least ionised component with log $\xi = 1.57 \pm 0.04$, but has the fastest outflow velocity $v_{out} = -1960 \pm 20$  km s\textsuperscript{-1}, and is responsible for the H-like and He-like \ion{O}{VII} and \ion{C}{VI}. In addition, component 3 produces the unresolved transition array (UTA) \citep{Behar2001} between 16 and 17 \AA, which can be seen in the bottom panel of Figure \ref{Model_WAs}. 

The H- and He-like absorption lines from \ion{C}{VI}, \ion{N}{VII}, \ion{O}{VII}, \ion{O}{VIII}, \ion{Ne}{IX} and \ion{Ne}{X} are fitted by all three absorption components. Components 1 and 2, with different outflow velocities, fit the lower energy part of the absorption lines, and due to its high blueshift velocity, component 3 produces a second trough in the absorption lines on the high energy side, or broadens the line. This can be seen in the absorption profile of the \ion{C}{VI} Ly$\alpha$ line, bottom panel in Figure \ref{Vel_Plots}, which indicates at what velocity each component produces an absorption feature of that particular ion. Due to the resolution of the RGS spectrum at lower energies it is clear that the \ion{C}{VI} Ly$\alpha$ absorption line contains two troughs from the three different WA components. 

\subsection{Emission line regions}
The emission lines in the spectrum are described by three \texttt{PION} components, with the best fit parameter values shown in Table \ref{Emission_Results}. The most dominant line produced by the first emission component (EM1) is the \ion{O}{VII} forbidden line at 22.4 \AA. The first emission component, EM1, also produces the \ion{N}{VI} forbidden line (30.0 \AA) and the \ion{C}{VI} Ly$\alpha$ line (34.4 \AA), in addition to the RRCs of \ion{C}{V} (32.1 \AA), \ion{C}{VI} (25.7 \AA), \ion{N}{VI} (22.8 \AA), and \ion{O}{VII} (17.0 \AA), with \ion{C}{V} being the most dominant RRC feature in the spectrum.

The strongest emission features produced by the second narrow emission component (EM2) are the \ion{C}{VI} Ly$\alpha$ and \ion{O}{VIII} Ly$\alpha$ (19.4 \AA) lines, along with the \ion{O}{VII} and \ion{Ne}{IX} (13.8 \AA) triplet lines. EM2 gives rise weakly to the \ion{Ne}{X} (12.3 \AA) and \ion{N}{VII} Ly$\alpha$ (25.2 \AA) lines and the \ion{O}{VII}, \ion{O}{VIII} (14.4 \AA), and \ion{C}{VI} RRCs; EM2 may also produce the \ion{Ne}{IX} RRC feature at around 10.5 \AA\, but the model could overestimate the spectrum here. We find that these emission features are emitted by EM2 at rest \citep[this was also found by][]{Mehdipour2018}. 

The third, broad, emission component (EM3) appears only to produce blueshifted forbidden lines of the \ion{O}{VII} and \ion{N}{VI} triplets, however the latter is not very evident in the spectrum. EM3 also produces a blueshifted \ion{C}{V} RRC.  

Figure \ref{Vel_Plots}, top panel, shows the velocity profile of the \ion{O}{VII} forbidden line with the emission components indicated by the purple dotted lines. The bottom panel shows the emission feature of the \ion{C}{VI} Ly$\alpha$, produced at zero km s\textsuperscript{-1}, with the position of the absorption components illustrated by the blue dotted lines.

\begin{table*}
	\caption{Best fit parameters for the three warm absorber components. The turbulent velocity of component 2 is coupled to that of component 1 as they have the same value when both are set free. We note the $\Delta C$ values, indicating the change in statistical fits produced by each component when introduced. Also shown are the ionic abundances of the photoionised and collisionally ionised plasmas throughout the nucleus of NGC 7469 (see Section \ref{Sec_Abund}).}
	\label{Absorption_Results}
	\centering
	\begin{tabular}{c | c c c c | c}
		\hline
		\hline
		Absorption & $N_{H}$  & log $\xi$ & $v_{turb}$ & $v_{out}$ & \multirow{2}{*}{$\Delta C$} \\
		Component & ($10^{24}$ m\textsuperscript{-2}) & ($10^{-9}$ Wm)& (km s\textsuperscript{-1}) & (km s\textsuperscript{-1}) & \\
		\hline
		1 & $10.0^{+0.5}_{-0.4}$ & $2.32 \pm 0.01$ & $35 \pm 2$ & $-630 \pm 20$ & 592 \Tstrut\Bstrut \\%
		2 & $52.0 \pm 2.2$ & 3.00 (f) & \ \tablefootmark{a} - & $-910^{+50}_{-30}$ & 105 \Tstrut\Bstrut \\%
		3 &$2.3 \pm 0.1$ & $1.57 \pm 0.04$ & $11 \pm 3$& $-1960 \pm 20$& 155 \Tstrut\Bstrut \\%
		\hline
			\hline
			
			\hline
		\multicolumn{6}{c}{Abundances\ \ \tablefootmark{b}} \Tstrut\Bstrut\\
		\hline
		\multicolumn{1}{c}{C} & N & Ne & Mg & \multicolumn{1}{c}{S} & Fe \Tstrut\Bstrut\\
			\multicolumn{1}{c}{$2.03^{+0.17}_{-0.08}$} & $1.11^{+0.20}_{-0.15}$ & $2.04 \pm 0.20$ &  $0.56^{+0.41}_{-0.13}$ &  \multicolumn{1}{c}{$2.80 \pm 0.40$} & $1.01 \pm 0.04$ \Tstrut\Bstrut \\
		\hline
	\end{tabular}\\
		\tablefoot{
		\tablefoottext{a}{Coupled to WA component 1.}
		\tablefoottext{b}{Abundances relative to Oxygen.} The $\log \xi$ parameter of WA component 2 is followed by a (f) because this value is fixed. }
\end{table*}

\begin{table*}
	\caption{Best fit parameters for the three emission components. In addition, we measure the emission measure (EM) of each component. We note the $\Delta C$ values, indicating the change in statistical fits produced by each component when introduced.}
	\label{Emission_Results}
	\centering
	\begin{tabular}{c | c c c c c c| c}
		\hline
		\hline
		Emission & $N_{H}$  & log $\xi$ & $v_{turb}$ & $v_{out}$ & $C_{cov} = $ & EM & \multirow{2}{*}{$\Delta C$} \\
		Component & ($10^{25}$ m\textsuperscript{-2}) & ($10^{-9}$ Wm)& (km s\textsuperscript{-1}) & (km s\textsuperscript{-1})  &$\Omega/4\pi$ & ($\times 10^{70}$ m\textsuperscript{-3})& \\
		\hline
		EM1 & $641 \pm 50$& $0.35^{+0.09}_{-0.01}$ & $50^{+140}_{-50}$ & $-660^{+110}_{-20}$ &$2.1 \pm 0.2 \times 10^{-4}$ & 13&230 \Tstrut\Bstrut \\
		EM2 & $42^{+7}_{-6}$& $1.55 \pm 0.08$ & $50^{+180}_{-30}$ & 0 (f) &$2.1 \pm 0.3 \times 10^{-3}$ & 1&34 \Tstrut\Bstrut \\
		EM3 &$787^{+130}_{-110}$ & $0.18^{+0.01}_{-0.07}$ &\tablefootmark{a} $1360^{+340}_{-270}$ & $-4460^{+200}_{-110}$ & $1.4 \pm 0.2 \times 10^{-4}$ & 15 &23  \Tstrut\Bstrut \\
		\hline
	\end{tabular}\\
	\tablefoot{
		\tablefoottext{a}{Broadening velocity ($v_b$) from the \texttt{VGAU} component which is coupled to the broad \texttt{PION} component.} The $v_{out}$ parameter of EM2 is followed by a (f) because this value is fixed.}         
\end{table*}

\subsection{Nuclear star burst ring}
The star burst region in NGC 7469 is found to have a CIE plasma temperature of $0.32 \pm 0.02 $ keV, with an outflow velocity of $-280^{+80}_{-70}$ km s\textsuperscript{-1} (Table \ref{CIE_Results}), consistent with previous results in this campaign \citep{Behar2017, Mehdipour2018}. In addition to the \ion{Fe}{XVII} emission lines at 15.3 and 17.4 \AA, the star burst region also emits the \ion{O}{VIII} Ly$\alpha$ line at around 19 \AA\ and the \ion{Ne}{IX} forbidden line at 13.9 \AA. The \texttt{CIE} component may also produce the \ion{O}{VIII} Ly$\beta$ and \ion{Ne}{X} Ly$\alpha$ lines.

\begin{table}
	\caption{Collisionally ionised plasma properties from the nuclear star burst region of NGC 7469.}
	\label{CIE_Results}
	\centering
	\begin{tabular}{c c c | c}
		\hline
	\hline
		EM  & T &$v_{out}$ & \multirow{2}{*}{$\Delta C$} \\ 
		($\times 10^{69}$ m\textsuperscript{-3}) & (keV) & (km s\textsuperscript{-1}) & \\
		\hline
		 $4.0^{+0.3}_{-0.29}$& $0.32 \pm 0.02$ & $-280^{+80}_{-70}$& 9 \Tstrut\Bstrut \\
		\hline
	\end{tabular}
\end{table}


\section{Characterising the emission and absorption regions}
 \label{EM_Characteristics}
In this section, we discuss the physical locations of the emission line region components, based on the results in Section \ref{Results} and Table \ref{Emission_Results}, and compare them with the distances of the WA. We also investigate the \ion{O}{VII} triplet, the most prominent feature in the RGS spectrum.
\subsection{Location of the emission line regions}
\label{EM_Dist}

The NLR distance in NGC 5548 was first calculated by \cite{Whewell2015}, and here we start by following their approach. We estimate the distance of the emission line regions within NGC 7469, assuming there is no absorption of the emission line region by the WA components (if we apply absorption from the WA components to each of the emission components we find no significant statistical improvement). We integrate the column density over the size of the emission line region, assuming constant ionisation parameter, as follows 
\begin{equation}
    N_H = \int^{r_{max}}_{r_{min}} n dr = \int^{r_{max}}_{r_{min}} \frac{L_{ion}}{\xi r^2} dr,
    \label{NH_EQ}
\end{equation}
where \textit{n} is substituted using Eq. \ref{XI_EQ}. For a single ionisation component, the integral on the right hand side of Eq. \ref{NH_EQ} yields
\begin{equation}
N_H = \frac{L_{ion}}{\xi} \left(\frac{1}{r_{min}} - \frac{1}{r_{max}}\right).
    \label{NH_EQ2}
\end{equation}
Assuming that the emission line region is a large, almost continuous cloud, then $r_{max} >> r_{min}$, which means the lower distance limit becomes
\begin{equation}
r_{min} = \frac{L_{ion}}{\xi N_H}.
\label{EQ_Whewell}
\end{equation}

Using the ionising luminosity of $L_{ion} = 1.39^{+0.02}_{-0.06} \times 10^{37}$ W, calculated from the SED, and the parameter values from Table \ref{Emission_Results}, the distances of the three emission components were calculated using Eq. \ref{EQ_Whewell}. The lower distance limits are shown in the left column of Table \ref{Emission_Distance_Results}, where it is clear EM3 is further away from the central source than EM1 and EM2. This is counter intuitive if we assume EM3 is part of the BLR, which should be closer to the black hole than the NLR (EM1 and EM2).

To overcome this, we use the definition of the column density which includes the volume filling factor $C_v$ ($N_H = C_v n \Delta r$) adopted by \cite{Blustin2005}. We integrate this column density over the length of the plasma in the emission line region, in the same way as for Eq. \ref{EQ_Whewell}, and obtain the new distance equation
\begin{equation}
    r_{min} = \frac{L_{ion} C_v}{N_H \xi}.
    \label{EM_Distances}
\end{equation}

In practice, however, the number density of the plasma is not constant and falls off as $r^{-2}$ (in the case where $C_v = 1$ (Eq. \ref{EQ_Whewell}), the hydrogen number density is assumed to be constant). If, however, we take two thin slices within our extended ($r_{max} >> r_{min}$) emitting plasma, one at $r_{min}$ and the other at $r_{max}$, assuming the number of hydrogen ions is the same in each shell, then the density in each shell is found to be proportional to $r^{-2}$, such that $n_{r_{min}} >> n_{r_{max}}$. Therefore, we can assume the hydrogen number density is negligible at distances larger than $r_{min}$, allowing us to directly substitute the hydrogen number density from Eq. \ref{XI_EQ} into Eq. \ref{NH_EQ} to obtain Eq. \ref{EM_Distances}. Further to this, the ionisation parameter in each emission line region is constant with respect to distance of the plasma\footnote{If $n = \frac{k}{r^2}$, where \textit{k} is the constant of proportionality, is substituted into Eq. \ref{XI_EQ}, the distances cancel, meaning the ionisation parameter $\xi$ has no dependence on the distance \textit{r}, so is constant throughout the plasma.}, such that all the ionisation occurs at $r_{min}$, where the density is largest.

There is very little information in the literature regarding the volume filling factor of the emission line regions, although the volume filling factor of the BLR is quoted between 0.001 and 0.01 \citep[e.g.][]{Osterbrock1991, Snedden1999}. For EM3, we set the filling factor at $C_v = 0.001$ \citep[see e.g.][and references within]{Snedden1999}, as we assume the broad line region is made up of multiple cloudlets. 



In order for the NLR to be further out from the central ionising source compared to the WA (assuming no absorption of the emission regions by the WA), we set $C_v$ to equal 0.1 for EM1 and EM2. 
Again, using the parameter values from Table \ref{Emission_Results} for the three emission components, the new emission line region lower limit distances were calculated and placed into the right column of Table \ref{Emission_Distance_Results}. We note that the uncertainties calculated in Table \ref{Emission_Distance_Results} do not include the large (and unknown) errors on the volume filling factor. Therefore, due to this, the uncertainties on the distance measurements are significantly greater than what is actually quoted. 

For each emission line region component, we calculate the emission measure (EM) to determine, from the contribution of each component, the line luminosities. The EM is given by 
\begin{equation}
EM = \frac{n_e}{n_H} 4 \pi C_{cov} \frac{L_{ion} N_H}{\xi}
\label{Em_Mes}
\end{equation}  
\citep[see e.g.][]{Mao5548, Psaradaki2018}, where $n_H$ and $n_e$ are the hydrogen and electron number densities, respectively, and the other parameters are calculated from the spectral modelling (see Table \ref{Emission_Results}). Eq. \ref{Em_Mes} is valid for any value of the volume filling factor and is not explicitly dependant on $C_v$. The calculated EM values (assuming $n_e = 1.2 n_H$ for fully ionised plasma) are displayed in Table \ref{Emission_Results}.

The latter method (using Eq. \ref{EM_Distances}) obtains a more reasonable result because Eq. \ref{EQ_Whewell} assumes, but does not explicitly state, a volume filling factor of 1. In NGC 5548, this means that the fraction of the total volume taken up by the single NLR was 1 \citep{Whewell2015}. In NGC 7469, with three emission components, this is unphysical as all the emission components cannot take up the whole volume. Therefore, we consider a volume filling factor such that each emission component only takes up a fraction of the total volume within NGC 7469.
\begin{table}
	\caption{Lower bound distance measurements of the emission line regions.}
	\label{Emission_Distance_Results}
	\centering
	\begin{tabular}{c | c c  }
		\hline
		\hline
		Emission & R & R \\
		Comp. &\ \ \ (pc)\ \ \tablefootmark{a} &\ \ \ (pc)\ \ \tablefootmark{b} \\
		\hline
		EM1  &$26^{+3}_{-7}$ & $2.62^{+0.31}_{-0.73}$  \Tstrut\Bstrut \\
		EM2  & $25^{+11}_{-8}$ & $2.52^{+1.05}_{-0.81}$\Tstrut\Bstrut \\
		EM3  & $30 \pm 10 $& $0.03 \pm 0.01$ \Tstrut\Bstrut \\
		\hline
	\end{tabular}\\
	\tablefoot{
		\tablefoottext{a}{All values are calculated with the volume filling factor of $C_v = 1$.}
		\tablefoottext{b}{EM1 and EM2 distances calculated using $C_v = 0.1$; EM3 distance calculated with $C_v = 0.001$.}}
\end{table}

\begin{figure}
	\centering
	\begin{subfigure}{1\linewidth}
		\includegraphics[width=\linewidth]{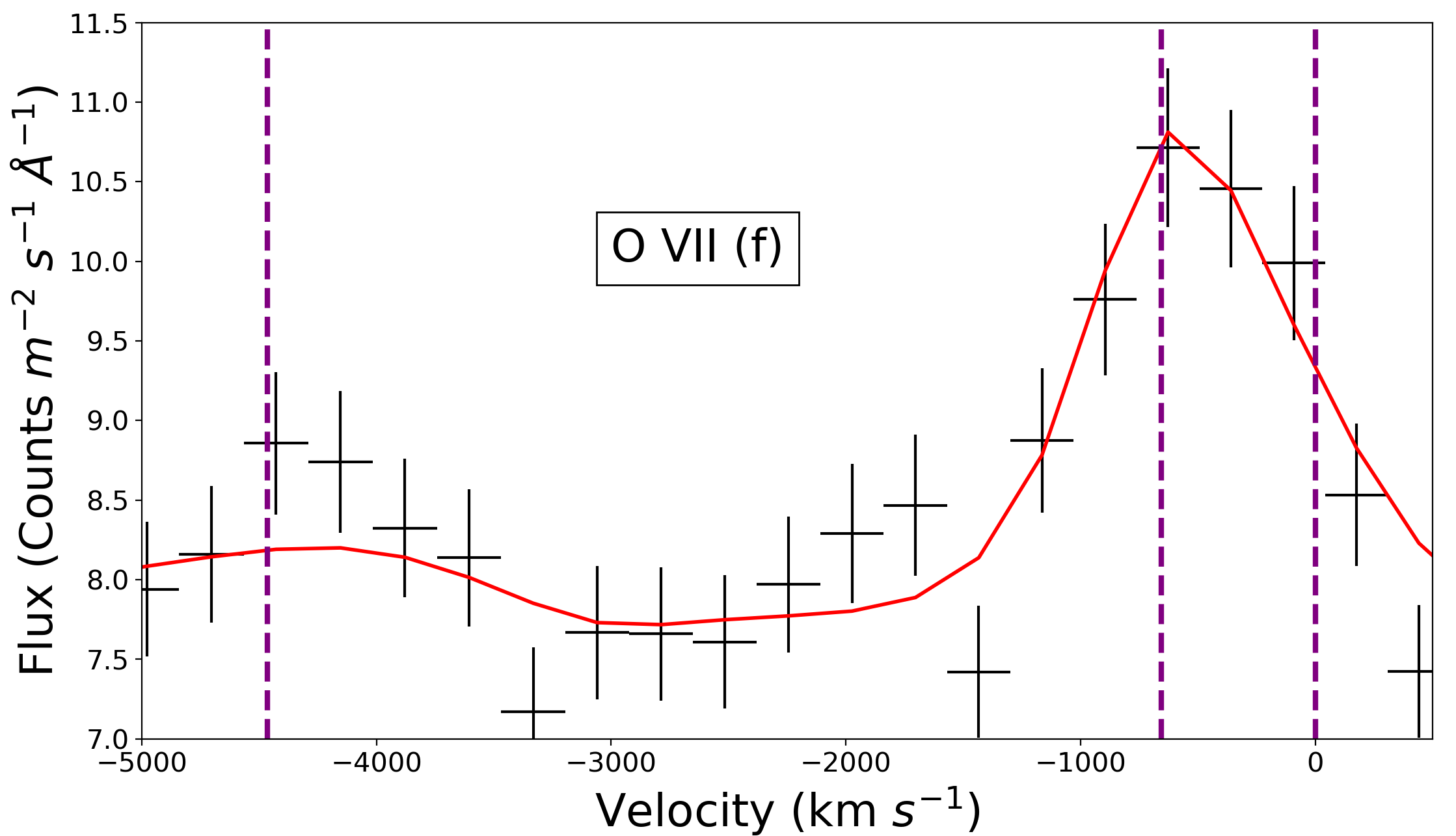}
	\end{subfigure}
	\begin{subfigure}{1\linewidth}
		\includegraphics[width=\linewidth]{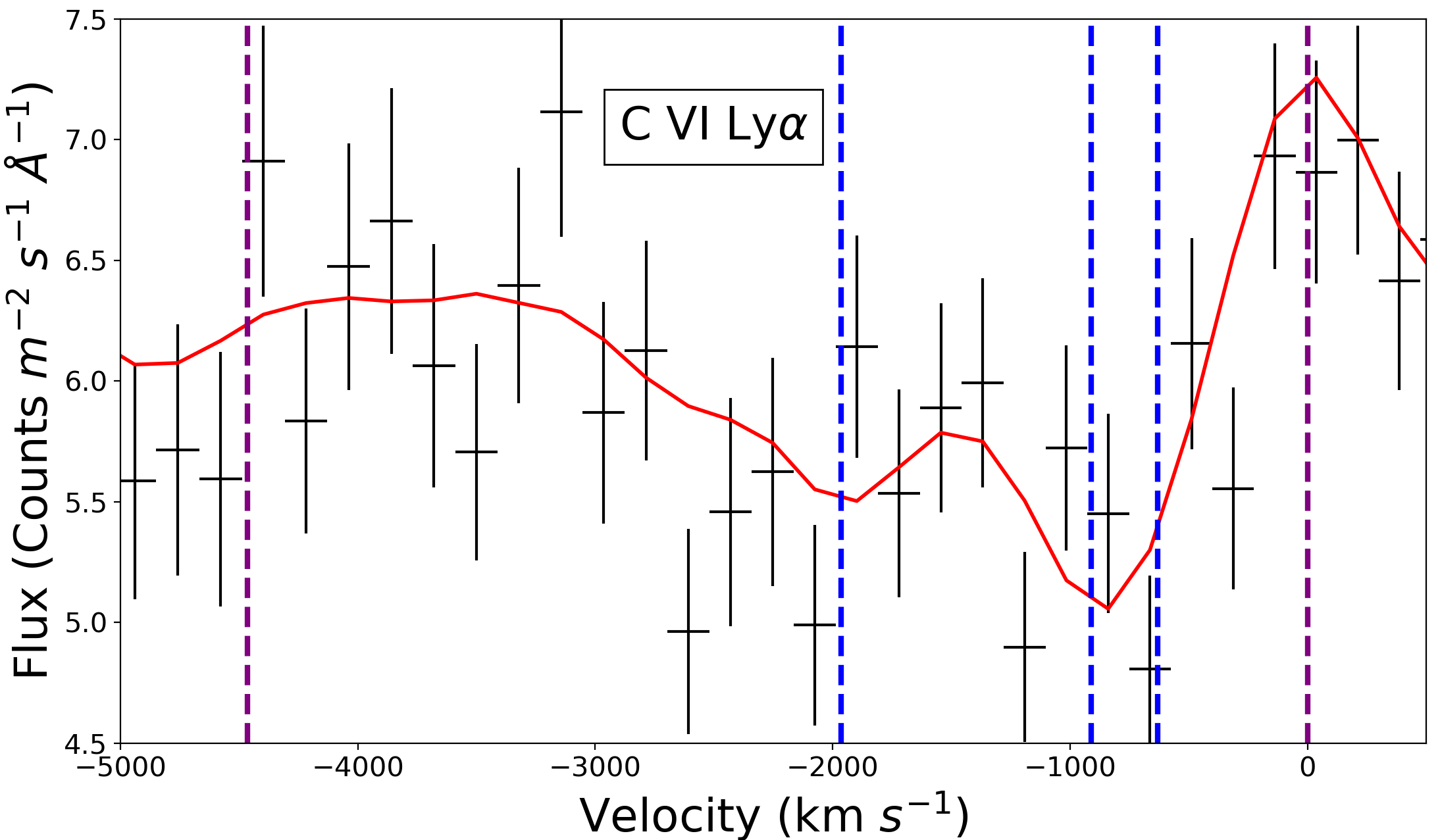}
	\end{subfigure}
	\caption{Velocity plots of the \ion{O}{VII} (f) line (top panel) and the \ion{C}{VI} Ly$\alpha$ line (bottom panel), illustrating each ion's velocity profile. \textit{Top Panel:} The \ion{O}{VII} forbidden line is produced by EM1 and EM2 at -657 and 0 km s\textsuperscript{-1}, respectively, as illustrated by the purple dotted lines showing the outflow velocities of the emission components. EM3 falls at the position of the intercombination line because the forbidden line is blueshifted to higher energies ($v_{out} -4464$ km s\textsuperscript{-1}). \textit{Bottom Panel:} The blue dotted lines indicate the velocities of the three WA components ($v_{out} =$ -631, -912 and -1964 km s\textsuperscript{-1}) and their contributions to the absorption. The \ion{C}{VI} Ly$\alpha$ line is produced at 0 km s\textsuperscript{-1}, which corresponds to EM2 (purple dotted line). We note that the widths of the emission lines in these two panels are determined fully by the resolution of the RGS instrument ($FWHM \sim 800$ km s\textsuperscript{-1} at 22 \AA\ and $FWHM \sim 600$ km s\textsuperscript{-1} at 34 \AA, for top and bottom panels respectively), and not by the turbulent velocities ($v_{turb}$) in Table \ref{Emission_Results}. }  
	\label{Vel_Plots}
\end{figure}

Although the distances between the narrow and broad emission components (from Table \ref{Emission_Distance_Results}, using Eq. \ref{EM_Distances}) are consistent with the overall picture of AGN, the distance of EM3 is still an order of magnitude larger compared to the distance of the optical BLR \citep[$r_{BLR} = 0.004$ pc;][]{Kollatschny2013}. This is because the ionisation parameter of EM3 is very small ($\log \xi = 0.18$), so Eq. \ref{EM_Distances} places this emission component further out than expected. This shows an inconsistency: if EM3 is a cloud in the BLR, then it would be bombarded with a large amount of X-ray flux, ionising the plasma greatly.

By comparing the optical BLR to EM3, the kinematics are similar: the FWHM of the $H\beta$ line in the optical is $2168 \pm 459$ km s\textsuperscript{-1} \citep{Kollatschny2013}, marginally consistent with the broadening velocity $v_b = 1360^{+340}_{-270}$  km s\textsuperscript{-1} of EM3 (an equivalent FWHM velocity of 3200 km s\textsuperscript{-1}); the radial velocity within the radius of the torus is -4521 km s\textsuperscript{-1} \citep{Suganuma2006}, which is again consistent with the EM3 outflow velocity $v_{out} = -4460^{+200}_{-110}$ km s\textsuperscript{-1}. Setting the outflow velocity as the escape velocity, the distance of EM3 results to be $R = \frac{2GM}{v^2_{out}} = 0.004 \pm 0.001$ pc. This value is closer to the black hole than the one found through the ionisation method (Eq. \ref{EM_Distances}), and is consistent with the optical distance, measured using the $H\beta$ line, of 4.5 ld, equivalent to 0.004 pc \citep{Kollatschny2013}.

This distance discrepancy, between the ionisation and kinematic methods is most likely due to the large uncertainties in the volume filling factor, which may be different up to many orders of magnitude. As stated before, these uncertainties are unknown, so the errors on the distance estimates cannot be quantified accurately.

\subsection{A thin shell approximation}
We note that our findings depend strongly on the assumption that the clouds emitting EM1 and EM2 are continuous and that they are located in a region with the outer radius much larger than the inner one. Alternatively, a thin shell model can be applied, given by $r_{max} = r_{min}(1 + \epsilon)$, where $\epsilon$ is a small number. From Eq. \ref{NH_EQ2}, introducing the volume filling factor $C_v$, and using the approximation $r_{min}(1 + \epsilon) \approx \frac{r_{min}}{1 - \epsilon}$\footnote{Using the Maclaurin series expansion.}, we obtain a distance to the emission components as follows 
	\begin{equation}
	r_{min} \approx \frac{L_{ion} C_v \epsilon}{N_H \xi}.    
	\label{Thin_Shell}
	\end{equation}
	This implies that the lower distance limit of the emission line region is proportional to $\epsilon$, meaning 
	the `true' lower distance limit measurements will be smaller than the values calculated in the right side of Table \ref{Emission_Distance_Results}. 

If we set $r_{min} \epsilon$ as the thickness of the shell $\Delta r$, then we can estimate the shell thickness, and thus obtain an updated distance estimate, based on the approximation above. This means that the Eq. \ref{Thin_Shell} becomes 
	\begin{equation}
	r_{min}^2 \approx \frac{L_{ion} C_v \Delta r}{N_H \xi}, 
	\label{shell_thickness}
	\end{equation}
	such that the distance ($r_{min}$) is proportional to the square root of the shell thickness. The shell thickness of EM1 can be estimated if we find the distance from the central source using the escape velocity as $v_{out} = -660$ km s\textsuperscript{-1} (see Table \ref{Emission_Results}), which results in a value of $r_{min} = 0.2$ pc. As the outflow velocity of EM2 is zero, we cannot estimate the shell thickness of this component. Rearranging Eq. \ref{shell_thickness}, substituting in $r_{min} = 0.2$ pc and using the parameters from Table \ref{Emission_Results}, we obtain a shell thickness of $\Delta r \simeq 0.01$ pc. Using the relation $\Delta r = r_{min} \epsilon$, $\epsilon = 0.05$. We then multiply $\epsilon$ by the distance of EM1 in the right side of Table \ref{Emission_Distance_Results} to get the new distance from the central black hole, assuming a thin shell geometry, which is $r_{min} = 0.13$ pc for EM1. However, these two values (0.2 and 0.13 pc) are inconsistent with each other, in a similar way to the kinematic and ionisation distances of EM3 (see end of previous section), most likely because of the chosen volume filling factor of EM1. 
	
	From these estimates, we can determine the volume filling factor, based on the kinematic distance of EM1. If we use $r_{min} = 0.2$ pc and $\epsilon = 0.05$, and Eq. \ref{Thin_Shell}, the volume filling factor is found to be $C_v \sim 0.15$ for EM1. We can also do this for EM3, as we know the kinematic distance too. Following the method above, the shell thickness of EM3 is $\Delta r = 4.2 \times 10^{-4}$, which means $\epsilon = 0.11$. From here, the derived volume filling factor of EM3 is $C_v = 9.6 \times 10^{-4}$. Both of these $C_v$ values for EM1 and EM3 are similar to the values assumed in Table \ref{Emission_Distance_Results}, right column. However, the $C_v$ and values derived here come from kinematic distance calculations, and not from the ionisation properties of the plasma.

This new distance places EM1 closer to the black hole than WA component 1, but within the upper and lower distance ranges of WA components 2 and 3. For EM1 to be further out from the black hole than all three WA components, we would require a volume filling factor larger than 0.1 (this is discussed further in Section \ref{AGN_Comp}).

\subsection{The O VII triplet}
\label{Oxygen_Triplet_Discussion}

The Oxygen He-like triplet is prominent in many AGN X-ray spectra \citep[e.g.][]{Whewell2015, Behar2017}. The velocities of the emission lines of the Oxygen triplet were found to differ in NGC 5548. The initial solution for this was that the emission lines were being absorbed by some of the WA components \citep{Whewell2015}. However, due to the contradictions in the implied geometry, \cite{Mao5548} considered multiple emission components, in much the same way as the WA is multiphased. \cite{Mao5548} found two narrow emission components explained this discrepancy as each had a different outflow velocity, with no influence on the geometry. In the case of NGC 3783, adding a broadened emission component significantly improved the fit by $\Delta C \sim 200$ \citep{Mao3783}. For this reason, we applied the same approach to NGC 7469.

Figure \ref{Oxygen_Triplet} displays the \ion{O}{VII} triplet between 21.7 and 22.6 \AA, with the best fit model (red line). The coloured lines show the contributions from each of the \texttt{PION} emission components one at a time, while the other two components are `switched off', that is to say, $N_H$ is set to zero so that no emission is included. EM2 (green line) fits the forbidden line, albeit not as strongly compared to EM1 (blue line), and also accounts for some of the resonance line. In Figure \ref{Oxygen_Triplet}, it is clear EM1 does not fit the resonance line. Initially, this was thought to be due to its large column density and optical depth, implying resonance scattering is the cause of no resonance emission. However, taking into account the very small covering fraction of EM1, we test to see if a smaller column density could be achieved, by increasing the covering fraction of EM1, allowing the resonance line to be fitted. Although increasing the covering fraction does decrease the column density, the resonance line is still not fitted by EM1, ruling out the possibility of resonant scattering being the cause. A degeneracy was found by \cite{DiGesu2017} between the column density and covering factor in their \texttt{PION} analysis of the Seyfert 1 galaxy 1H 0419-577. Further to this, we refit these two parameters ($N_H$ and $C_{cov}$), and the best fit is again obtained for the parameters in Table \ref{Emission_Results}. Even a local fit of just the \ion{O}{VII} triplet (between 21.7 and 22.6 \AA) requires EM1 to have similar parameter values to those in Table \ref{Emission_Results}; again the resonance line is unaccounted for. This, therefore, suggests that EM1 is fairly compact, with large column density and small ionisation, similar to EM3. However, we then have the problem of the narrow component EM1 having similar ionisation properties to those of the broad emission component (see the following paragraph), while EM1 does not contribute to the resonance line.

 Although both EM1 and EM2 fit the majority of the emission features in the rest of the spectrum, the intercombination line is not accounted for by either of them. (The small peak from EM2 at 22.1 - 22.2 \AA\ is not very conclusive in Figure \ref{Oxygen_Triplet}.) This is where the broad emission component came in.

EM3 appears to fill in the intercombination line in the spectrum (see Figure \ref{Oxygen_Triplet}, labelled with the purple \textit{f}). However, due to the high blueshift velocity of the broad component, $v_{out} = -4460^{+200}_{-110}$ km s\textsuperscript{-1}, the peak of the forbidden line (in the observed frame) is shifted towards shorter wavelengths, so falls at the position where the intercombination line would be expected. This is evident in the velocity profile of \ion{O}{VII} (f) in the top panel of Figure \ref{Vel_Plots} (see also the purple line in Figure \ref{Oxygen_Triplet}); the intercombination line is actually filled in coincidentally. If EM3 is part of the BLR, then we would expect a strong resonance line to be emitted, due to the high density plasma in the BLR, blueshifted to higher energies. However, the forbidden line is the strongest feature produced by EM3, making the interpretation of the broad emission component doubtful. This is also supported by the ionisation parameter which is very small (Table \ref{Emission_Results}). Figure \ref{EM3_Density} displays the \ion{O}{VII} concentration as a function of $\xi$ for EM3 as derived using the \texttt{PION} code. The ionisation parameter value for EM3 (log $\xi = 0.18$; red line in Figure \ref{EM3_Density}), shows the ionisation level is far from the peak for the \ion{O}{VII} ion, implying this region is not ionised enough to produce a substantial amount of forbidden line. Therefore, EM3 only statistically improves the fit, and is likely to represent an unphysical, ad hoc solution.

\subsection{The missing O VII intercombination line} 
After ruling out the physical validity of EM3 to explain the residuals between 22.0 and 22.2 \AA, where the \ion{O}{VII} intercombination emission line is, we investigate why EM1 and EM2 do not produce the intercombination line. One reason why the intercombination line is weak (or non-existent), could be due to a low plasma density, which would also enhance the forbidden line. 
For high density plasma, the excitation from the upper level of the forbidden line occurs from electron collisions, and this would therefore result in a stronger intercombination line, which is not seen here. With the \texttt{PION} model, we tried to calculate the densities of the narrow emission components, however we were unable to constrain the values \citep[see e.g. Figure 9 in][]{Porquet2010}. Instead, upper limits to the densities in EM1 and EM2 were calculated using Eq. \ref{XI_EQ}, where the ionisation parameters are from Table \ref{Emission_Results}, the lower limit distances are from the right side of Table \ref{Emission_Distance_Results} and $L_{ion} = 1.39 \times 10^{37}$ W. The density upper limits were found to be $n_1 \leq 1 \times 10^{12}$ m\textsuperscript{-3} for EM1 and $n_2 \leq 7 \times 10^{11}$ m\textsuperscript{-3} for EM2; these values are consistent with the densities found in the NLR \citep[$n_{NLR} \sim 10^{12}$ m\textsuperscript{-3}; e.g.][]{Netzer1990}. This relatively low density is consistent with a strong forbidden line.

\begin{figure*}
	\centering
	\includegraphics[width=1\linewidth]{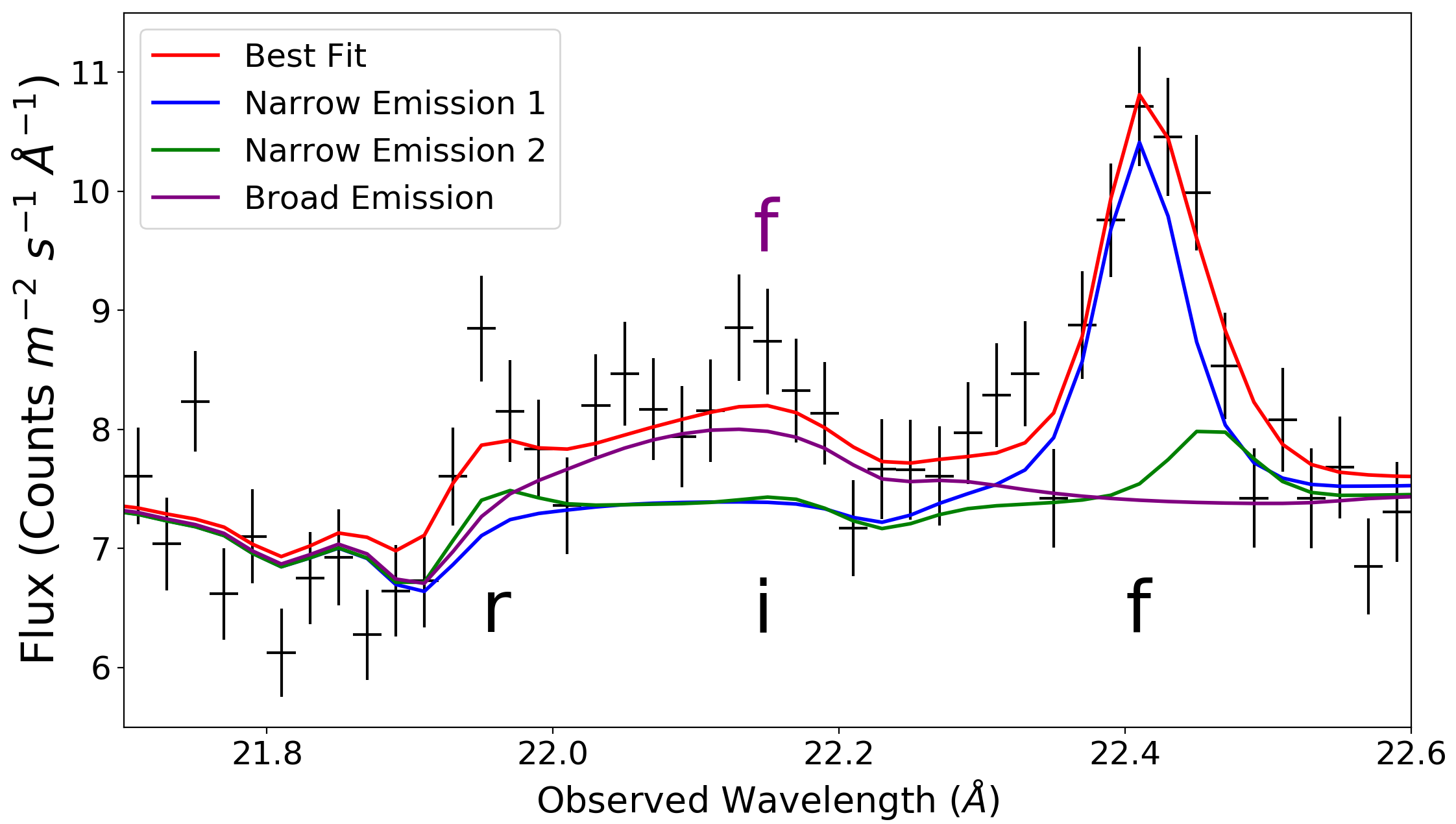}
	\caption{Plot of the \ion{O}{VII} He-like triplet in the observed frame of NGC 7469, with the three emission lines labelled in black (resonance, intercombination and forbidden for r, i and f, respectively). The spectrum (black crosses) is over imposed with the best fit model (red line). Two narrow components (EM1 and EM2) are shown by the blue and green lines, respectively, and a broad component (EM3) is plotted with the purple line. 
		EM1 fits the forbidden line, EM2 fits the resonance line, as well as the forbidden line, and EM3 produces emission at location of the intercombination line, which is actually a broadened forbidden line with a large blueshifted velocity (as labelled with the purple `f').}
	\label{Oxygen_Triplet}
\end{figure*}

\begin{figure}
\centering
\includegraphics[width=1\linewidth]{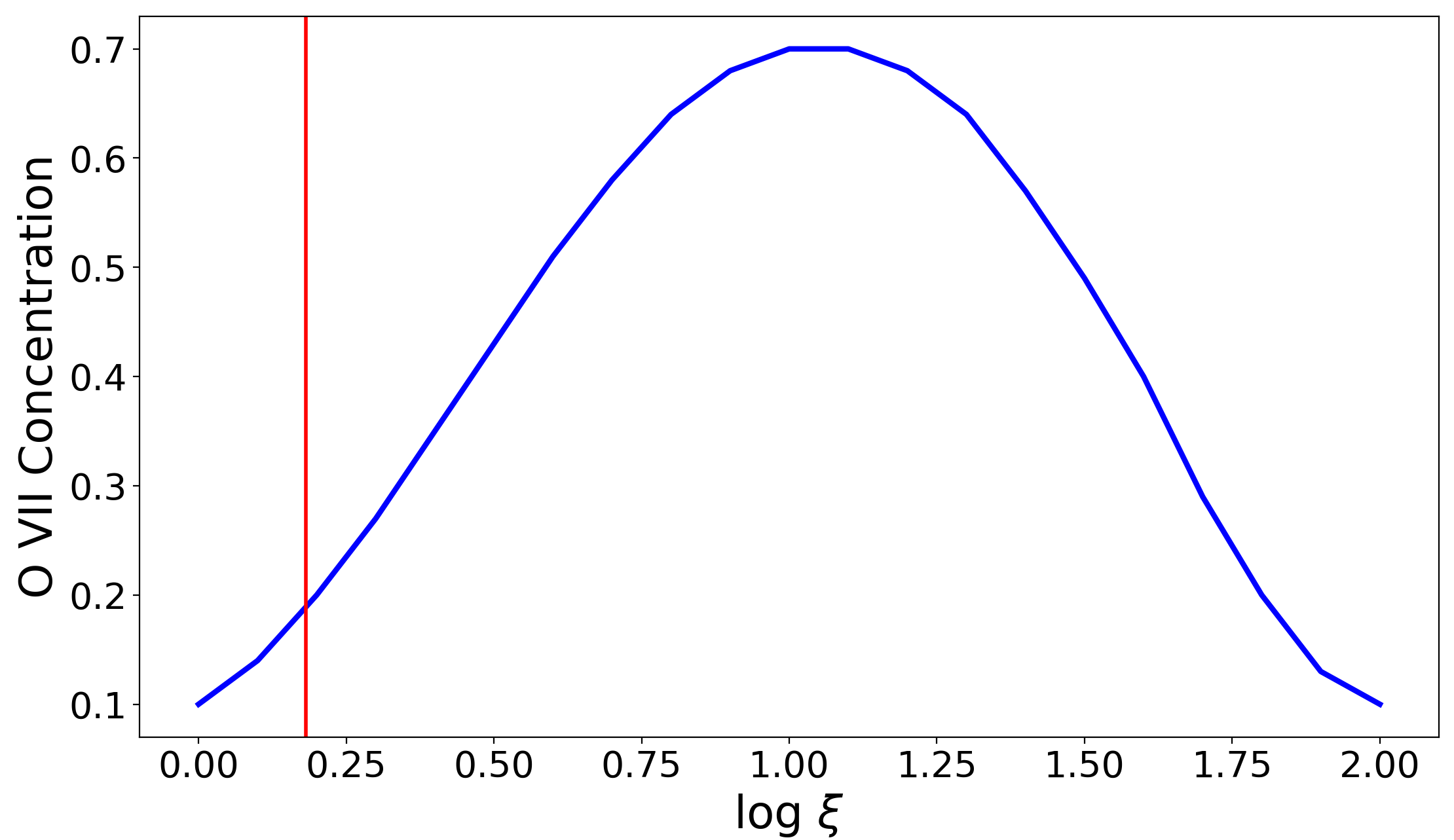}
\caption{Ionic concentration of \ion{O}{VII} as a function of ionisation parameter $\xi$ (blue), calculated using \texttt{PION}. The red line shows the value of $\xi$ found for EM3 through photoionisation modelling, which is much lower than the peak \ion{O}{VII} ion concentration, indicating that this region is not ionised enough to produce a substantial forbidden line.}
\label{EM3_Density}
\end{figure}

An alternative solution to the lack of intercombination emission could be Li-like absorption \citep{MehdKaa2015}. This is where the \ion{O}{VI} ion self-absorbs the intercombination emission line of the \ion{O}{VII} triplet at around 22 \AA\ (in our reference frame; Figures \ref{PION_SPEC_2} and \ref{Oxygen_Triplet}). This can be seen in the rest frame in Figure 1 of \cite{MehdKaa2015}.

To test this out, we use the \texttt{SPEX} model \texttt{SLAB}, which applies a single absorption phase. By keeping all the best fit parameters fixed, we multiplied a \texttt{SLAB} component to the two narrow emission line components and fitted the \ion{O}{VI} column density, with inital value of $\log N_{\ion{O}{VI}} = 22$ m\textsuperscript{-2} \citep{MehdKaa2015}. However, there is no significant change in C-statistic, nor any absorption features in the spectrum that would be produced by \ion{O}{VI}. One problem is that \texttt{SLAB} models only the foreground absorption, and not the self-absorption in the plasma; the transmission calculations within the plasma are more complex. In conclusion, it is still unclear how to explain the presence of emission at the position of the \ion{O}{VII} intercombination line given the emission components which best fit the rest of the spectrum of NGC 7469.

\subsection{Searching for LAOR broadened emission}
\label{Laor_Sec}
In the NGC 7469 spectrum, both in Figure \ref{PION_SPEC_2} and in \cite{Behar2017}, there are some unaccounted for emission features at 32.6 and 33.4 \AA\ (in our reference frame). We investigate these features using a \texttt{LAOR} component \citep{Laor1991}, which accounts for relativistically redshifted and broadened emission from the accretion disc; a similar method was used by \cite{Graziella2001} on two AGN with peculiar spectra. Reflection at low energies is blurred by relativistic effects from the accretion disc \citep[e.g.][]{Blustin_Fabian2009}, which may be an explanation for the soft excess, found from soft X-ray time lags \cite[see e.g.][and references within]{DeMarco2018}.  To model these emission features, we swap the \texttt{VGAU} component for a \texttt{LAOR} component, coupled this to the third \texttt{PION} emission component and fitted, fixing all the best fit parameters except for EM3. 

This new fit removes the intercombination line of the \ion{O}{VII} triplet modelled by EM3 (increasing the residuals at 22.2 \AA), but reduces the residuals at 23.5 \AA, where the Galaxy \ion{O}{I} absorption is present. However, the unaccounted for emission features at 32.6 and 33.4 \AA\ are not fitted. The fit only improves by $\Delta C = 14$, too small to determine any significant change to the spectrum, therefore, \texttt{LAOR} emission is not investigated further. 

\subsection{The WA location}
Although this paper focusses on the emission line region within NGC 7469, the absorption features from the WA had to be reanalysed as well. Therefore, from the best fit parameters of Table \ref{Absorption_Results} that describe the WA components, we were able to estimate their locations. The location of the WAs in many AGN is still debated, and their origins are still unclear; most researches favour the launching site to be at the torus \cite[e.g.][]{Krolik2001, Blustin2005}. To estimate the WA distances from the black hole, we used the lower and upper distance limits, following \cite{Blustin2005}, given by 
\begin{equation}
R \geq \frac{2GM_{BH}}{v^{2}_{out}},
\label{WA_Lower}
\end{equation}
where $M_{BH}$ is the black hole mass \cite[$M_{BH} = 10^{7} M_{\sun}$;][]{Peterson2014}, using the expression for the escape velocity from the gravitational potential of the central SMBH, and
\begin{equation}
R \leq \frac{L_{ion}C_{v}}{\xi N_{H}},
\label{WA_Upper}
\end{equation}
where $C_v$ is the volume filling factor. The upper distance limit (Eq. \ref{WA_Upper}) is derived from $N_H \approx n C_v \Delta r$, where \textit{n} can be substituted for using Eq. \ref{XI_EQ}, and we assume the size of the WA ($\Delta r$) to be equal to or smaller than its distance ($R$) to the SMBH \citep[$\frac{\Delta r}{R} \leq 1$;][]{Blustin2005}. To calculate $C_v$, we use 
\begin{equation}
C_v = \frac{(\dot{P}_{abs} + \dot{P}_{scat})\xi}{1.23 m_p L_{ion} v^2 \Omega},
\label{Vol_Filling}
\end{equation}
where $\Omega = 1.6$, generated by assuming a quarter of near AGN are type 1, with an outflow covering factor of at least 0.5 \citep{Blustin2005}. The derivation of $C_v$ assumes that the momentum of the outflowing wind is related to the momentum of radiation being absorbed ($P_{abs}$) plus the momentum of the scattered radiation ($P_{scat}$), which depends on the size of the WA, related to $C_v$. Here, $\dot{P}_{abs}$ and $\dot{P}_{scat}$ are
\begin{equation}
\dot{P}_{abs} = \frac{L_{abs}}{c};\ \dot{P}_{scat} = \frac{L_{ion}}{c} (1 - e^{-\tau_T});\ \tau_T = \sigma_T N_H, 
\end{equation}
where $\tau_T$ is the optical depth, $\sigma_T$ is the Thompson scattering cross-section and $L_{abs}$ is the absorbed luminosity. To calculate $L_{abs}$, we assume that each WA component is separate to each other and that there is no overlap between them in our line of sight to the central engine. However, this is unlikely to be the case, and when we measure $L_{abs}$ for all three components combined, the absorbed luminosity is larger than the three individual values, implying there is some overlap. Using the WA parameter values from Table \ref{Absorption_Results} and Eqs. \ref{WA_Lower}, \ref{WA_Upper}, and \ref{Vol_Filling}, the values of $\dot{P}_{abs}$, $\dot{P}_{scat}$, $C_v$, $R_{min}$, and $R_{max}$ are calculated (see Table \ref{Blustin_Distances}). 

Here, the torus distance (due to dust sublimation) is estimated using 
\begin{equation}
R_{torus} \approx 10^{-2} \sqrt{L_{ion}},
\label{Torus_Radius}
\end{equation}
where $L_{ion} = 1.39^{+0.02}_{-0.06} \times 10^{37}$ W is calculated from the SED, and $R_{torus}$ is measured in metres and $L_{ion}$ is in Watts; this equation is an amendment to Eq. 1 from \cite{Ashton2006} where $R_{torus}$ was measured in cm and $L_{ion}$ was measured in erg s\textsuperscript{-1}. From this, the torus is at a distance of $1.21^{+0.02}_{-0.05}$ pc. The escape velocity from the black hole at the torus is then $\sim 189$ km s\textsuperscript{-1}, which is far less than the outflow velocities of the three WA components. Therefore, these WA components can originate, or be located, closer than the torus.   


\begin{table*}
	\caption{Change of momentum (due to absorption and scattering), volume filling factor ($C_v$) and distance measurements of each WA component, along with mass outflow rates, kinetic luminosities and their fractions with respect to the bolometric luminosity \citep[$L_{bol} = 2.5 \times 10^{37}$ W;][]{Petrucci2004}.}
	\label{Blustin_Distances}
	\centering
	\begin{tabular}{c | c c c c c c c c}
		\hline
		\hline
		WA  & $\dot{P}_{abs}$  & $\dot{P}_{scat}$  &  $C_v$  & $R_{min}$  & $R_{max}$  & $\dot{M}$  & log $L_{K}$ & $L_{K}$ / $L_{bol}$ \Tstrut\Bstrut\\
		Comp.& (Ws m\textsuperscript{-1}) & (Ws m\textsuperscript{-1}) & & (pc) & (pc) & ($M_{\sun}$ yr\textsuperscript{-1}) & (W) & \\
		\hline
		1 & $7.3 \times 10^{26} $& $3.1 \times 10^{25}$ &0.0087 & $0.22 \pm 0.01$ & $1.88^{+0.42}_{-0.21}$ & 0.019 & 32.4 & $1.0 \times 10^{-5}$ \Tstrut\Bstrut\\
		2 &$9.3 \times 10^{26}$ &$1.6 \times 10^{26}$ & 0.0290 & $0.10^{+0.01}_{-0.02}$& $0.25 \pm 0.02$& 0.019 & 32.8 & $2.5 \times 10^{-5}$ \\
		3 & $5.3 \times 10^{26}$ &$7.0 \times 10^{24}$ & 0.0001& $0.02 \pm 0.01$& $0.60^{+0.13}_{-0.16}$& 0.004 & 32.6 & $1.6 \times 10^{-5}$\\
		\hline
	\end{tabular}
\end{table*}

In addition, the mass outflow rate and kinematic luminosities of each WA component were calculated. This gave us an insight on the amount of matter being carried away by the AGN outflow, and the luminosities of these individual components. The mass outflow rate ($\dot{M}_{out}$) and kinetic luminosity ($L_K$) are given by \citep{Blustin2005}
\begin{equation}
\dot{M}_{out} = \frac{1.23m_p L_{ion} C_{v}v_{out}\Omega}{\xi},
\label{MassRate}
\end{equation}
and 
\begin{equation}
L_K = \frac{\dot{M}_{out} v^2_{out}}{2} = \frac{1.23m_p L_{ion} C_{v}v^3_{out}\Omega}{2 \xi},
\end{equation}
where $m_p$ is the proton mass, $\Omega = 1.6$ is the solid angle, $C_v$ is from Table \ref{Blustin_Distances}, and $L_{ion} = 1.39 \times 10^{37}$ W is the ionising luminosity from the central source. 

The WA component values for the mass outflow rate and kinematic luminosities are displayed in Table \ref{Blustin_Distances}. The total mass outflow rate of the three components is 0.042 $M_{\sun}$ yr\textsuperscript{-1}, which is somewhat less than the total of 0.052 $M_{\sun}$ yr\textsuperscript{-1} for the three components in archive data \citep{Blustin2007}.

We compare the mass outflow rate (Eq. \ref{MassRate}) with the mass accretion rate, $\dot{M}_{acc} = \frac{L_{ion}}{\eta c^2}$, and find $\dot{M}_{out} \sim 2 \dot{M}_{acc}$ (for all three WA components summed together). Using the relation $\dot{M}v = \frac{L_{Edd}}{c}$, where $L_{Edd} = \eta \dot{M}_{acc} c^2$ is the Eddington luminosity, and assuming $\eta = 0.1$ and $v \sim 3000$ km s\textsuperscript{-1} (a smaller velocity would give a larger mass outflow rate), the ratio $\dot{M}_{out} \simeq 10 \dot{M}_{acc}$ is found. This difference indicates that there could be some other driving mechanisms at work, in addition to the spherically symmetric radiation outflow assumed here, such as a thermally-driven \citep[e.g.][]{Krolik2001} or magnetohydrodynamic \citep[e.g.][]{Fukumura2010} outflow.
	

The $L_K$ values are consistent between the two epochs \citep[31.7 to 32.7 W from][]{Blustin2007}. We also measure $L_K$ as a fraction of the bolometric luminosity $L_{bol} = 2.5 \times 10^{37}$ W \citep{Petrucci2004}. The values for this fraction are shown in Table \ref{Blustin_Distances} and are between $1 - 3 \times 10^{-3}$\% of $L_{bol}$, consistent with previous analysis \citep{Blustin2007}. The $L_K$ of the UV outflowing material is negligible compared to the X-ray components \citep{Aravprep}, so does not need to be included here.

\section{Discussion}
\label{Discussion}
\subsection{Physical structure of NGC 7469}

\begin{figure*}
	\centering
	\includegraphics[width=0.8\linewidth]{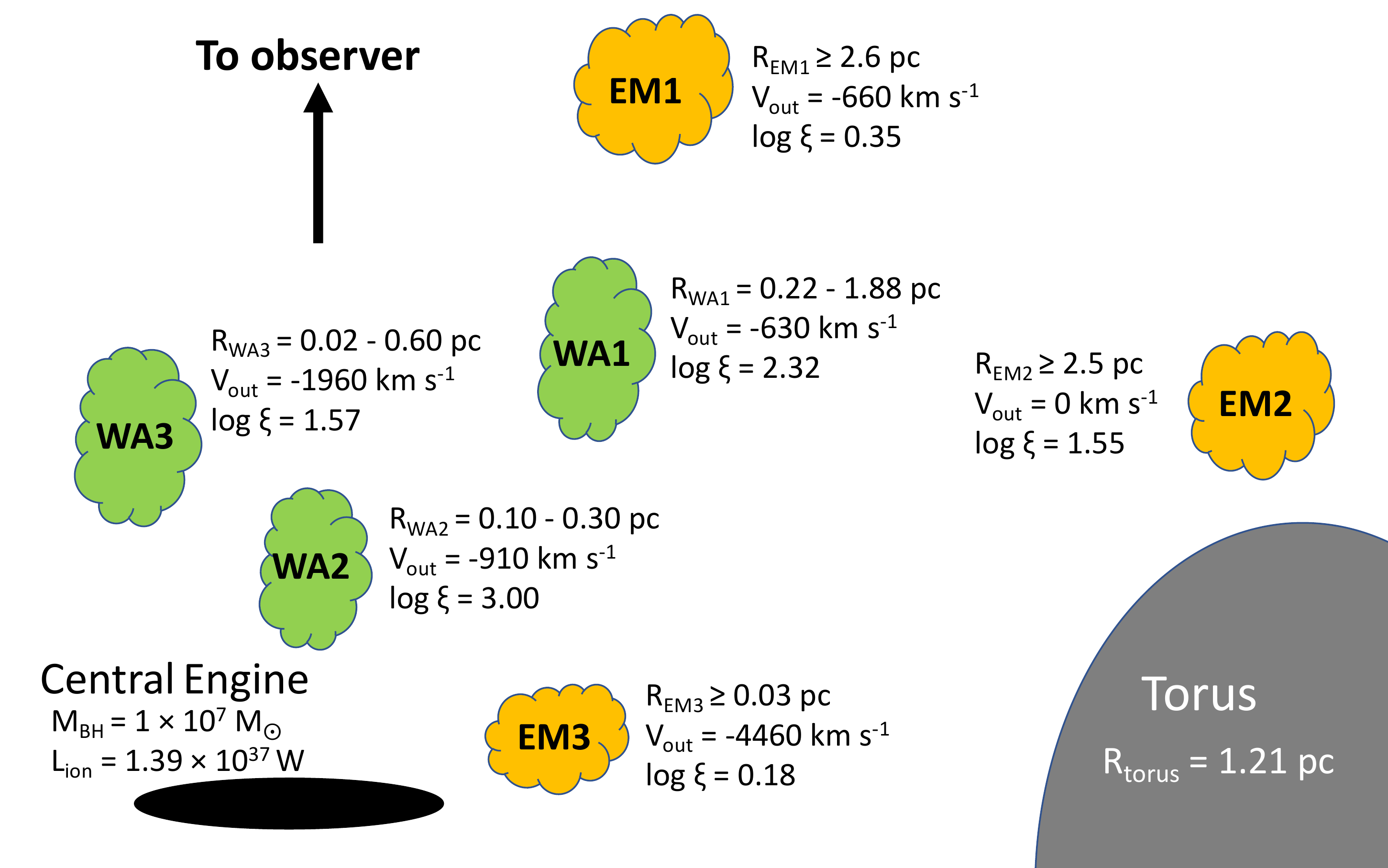}
	\caption{Schematic display of the possible locations of the WAs (green) and emission line regions (orange) with respect to the central black hole. This graphic is not intended to show the exact locations of the photoionised plasma clouds, but can give insight into studying the physical scenarios. Next to each plasma cloud are the distances, ionisations and outflow velocities, depicted for an easy comparison. (The distances for the emission line regions are the lower limits, as obtained from Eq. \ref{EM_Distances}.) As a comparison to the X-ray broad emission component (EM3), the optical BLR is at a distance of 0.004 pc in NGC 7469 \citep{Kollatschny2013}.}
	\label{AGN_Structure}
\end{figure*}

Figure \ref{AGN_Structure} displays a schematic to demonstrate the structure and possible locations of the emission line regions and WA components in NGC 7469 with respect to each other, the torus, and the central engine. This diagram is based upon the distances of each photoionised plasma phase from the nuclear black hole, which aids in the study of the physical structures and possible scenarios.

The ionisation parameters of both EM2 and WA component 3 are $\log \xi = 1.6$, suggesting that they could be part of the same, extended photoionised plasma \citep[e.g.][]{Kinkhabwala2002, Blustin2003, Behar2003, Behar2017}. However, there is not enough evidence to conclusively deduce this, as the location of EM2 is very different compared to WA component three. The outflow velocity of EM2 is fixed at rest in the best fit model (Table \ref{Emission_Results}), suggesting that it does not have any radial velocity (or it is at least small relative to the other emission line components). Therefore, in Figure \ref{AGN_Structure}, EM2 is placed as close to the plane of the torus as possible, but such that it can still receive the full ionising continuum from the central source. Consequently, although these two components share the same ionisation parameter, it is unlikely that they are part of the same, extended wind, because of their varying outflow velocities and the location of EM2 with respect to the WA components. 

On the other hand, both EM1 and WA component 1 have similar outflow velocities, but different ionisation parameters. This suggests that the emission line region could also be part of the outflowing wind. If these two components were part of the same outflowing wind then EM1 may have a smaller ionisation parameter because WA component 1 is obscuring the ionising flux getting to it. However, whether the plasma emits or absorbs ultimately depends on the underlying densities of the plasma regions.

From Figure \ref{AGN_Structure}, it looks like EM3 ($r = 0.03$ pc; $v_{out} \sim 4500$ km s\textsuperscript{-1}) could catch up with WA component 2 ($r_{min} = 0.1$; $v_{out} \sim 1000$ km s\textsuperscript{-1}) in just over 30 years. However, this is not a secure prediction for many reasons. Firstly, there is a difference (by an order of magnitude) between the calculated distances of EM3, depending on if we use the ionisation parameter (Eq. \ref{EM_Distances}) or the kinematics (Eq. \ref{WA_Lower}). Secondly, a large uncertainty in the volume filling factor means the distance could be very different to the value quoted in the right side of Table \ref{Emission_Distance_Results}; in addition, a $C_v < 0.001$ could mean the ionisation distance is similar to the kinematic distance. Finally, the warm absorber distances calculated here are at least 10 times smaller than the distances from variability arguments \citep{Mehdipour2018}, meaning WA component 2 could be much further away from EM3, thus implying it is very uncertain whether EM3 could catch up to WA component 2.

Therefore, we cannot conclusively say that some of the plasma clouds are made up of both absorption and emission regions in the nucleus of NGC 7469. Furthermore, it becomes obvious, when comparing the upper distances of the three WA components in Table \ref{Blustin_Distances} to the distances of the components found by \cite{Mehdipour2018} ($2 < r < 31$, $12 < r < 29$, $r< 31$, and $r< 80$ pc for the four WA components), and the derived lower distance limits from \cite{Peretz2018} ($r > 12 - 31$ pc), that the different analysis methods adopted can lead to different results, which therefore gives alternatives to the schematic in Figure \ref{AGN_Structure}.  \cite{Mehdipour2018} used the variability technique and a recombination time of 13 years to determine how the WA winds changed to the shape of the SED. Alternatively, we have assumed a thin shell for each WA component, where all of the ionisation occurs \citep[see e.g.][]{Blustin2005} because we are only using data from 2015, and the spectrum does not show variability \citep{Behar2017}. Therefore, it is unsurprising that, using the method from \cite{Blustin2005}, the distances found here are consistent with previous findings of the WAs within NGC 7469 \citep[0.1 - 1.6 pc; 0.012 - 1.7 pc,][respectively]{Blustin2005, Blustin2007}. The ability to constrain distance estimates from variability arguments naturally depends on the timescales covered by the data; given the long intervals between observations, the derived distances are generally (large) lower limits. Here we try to find a way to reconcile our results with the expectations based on the standard model of AGN by adjusting the $C_v$ parameter.
\subsection{Comparison with other AGN} 
\label{AGN_Comp}
In NGC 5548, the X-ray narrow line region was found to be located at a distance of 13.9 pc \citep{Whewell2015}, but the volume filling factor was set at 1 by these authors (see Section \ref{EM_Dist}). If $C_v = 0.1$, then the narrow line region would be situated at 1.39 pc, which is consistent with the optical narrow line region, found between 1 to 3 pc \citep{Peterson2013}. However, this is in variance with the overall picture of AGN in NGC 5548 if the WAs are located between 5 to 10 pc \citep{Ebrero2016}, as the narrow emission line regions are generally expected to be found further out than the WAs \citep[e.g.][]{Mao5548}. In NGC 7469, the WA distances derived from variability arguments have a lower limit of $r > 12 - 31$ pc \citep{Peretz2018, Mehdipour2018}, meaning that our current narrow emission line distances ($\sim 2.5$ pc, right column in Table \ref{Emission_Distance_Results}) would be too small. If the volume filling factor of the narrow line regions were 1 (Eq. \ref{EQ_Whewell}), then the lower limit distance values for EM1 and EM2 would become 25 and 26 pc, respectively (see left side of Table \ref{Emission_Distance_Results}), taking these distances to be of the same order as the WA components. Using the emission parameters from the Chandra data \citep[Table 3 in][]{Mehdipour2018}, we calculate the minimum distances of the narrow emission line regions at 2.4 and 22.3 pc (with $C_v = 0.1$), consistent with the results in the right column of Table \ref{Emission_Distance_Results}. Therefore, both sets of results (RGS and Chandra) can be consistent with the narrow emission line regions being further out that the WA components \citep{Peretz2018, Mehdipour2018} if the filling factor is between 0.1 and 1 for the emission line regions. The range in $C_v$ allows for this discrepancy of the NLR distances in NGC 7469, and in NGC 5548, to be overcome. 

We have also calculated the distances of the emission line regions within NGC 5548 and NGC 3783, both analysed with \texttt{PION}, by \cite{Mao5548} and \cite{Mao3783}, respectively, using Eq. \ref{EM_Distances} (for the narrow emission components $C_v = 0.1$, and for the broad components $C_v = 0.001$). The lower distance limits of the emission line regions within these AGN are displayed in Table \ref{Dist_5547_3783}. For NGC 5548, we find the lower distance limits of the narrow emission line regions 
for Model D and 
Model T, which has the addition of the broad emission component \citep[see model explanations in][]{Mao5548}. 
The results for Model D are consistent with the NLR distance of 13.9 pc from \cite{Whewell2015}. The large, measurable, difference in distances between NGC 7469 and NGC 5548 is due to the column densities in NGC 7469 being an order of magnitude larger than the equivalent values in NGC 5548. However, as the SMBH mass in NGC 5548 is about 7 times more massive than the SMBH in NGC 7469 \citep{Bentz2007, Pancoast2014}, it is expected that the emission line regions in NGC 5548 are found further out from the black hole (the distance is proportional to the black hole mass, for example Eq. \ref{WA_Lower}).

In NGC 3783, we derive the lower distance limits of the emission line regions from the results found by \cite{Mao3783}, from three observations: 2000/01, and 11 and 21 December 2016 (see Table \ref{Dist_5547_3783}). 
From these results, it appears extremely unrealistic to have component 2 move over 10 pc in the 15 years or so between observations; the broad component has not moved within this time frame. On the other hand, what is striking here is that the first narrow emission component has a distance comparable to that of the broad component. This could be because the ionisation parameter of emission component 1 is greater compared to the other two components, and the volume filling factor of the broad component is set at 0.001. Again, this is at variance with the generally accepted view of the central region of AGN, whereby the NLR is further away from the central engine than the BLR. Once again, the volume filling factor for the narrow emission line region needs to be between 0.1 and 1, such that the upper limit allows for the narrow emission component 1 in NGC 3783 to be 10 times further away than the BLR. Alternatively, it is possible that the volume filling factor of the broad emission component is less than 0.001, similar to why EM3 has an ionisation distance an order of magnitude larger than the kinematic distance (see Section \ref{EM_Dist}).

Further investigation of the filling factor value for the narrow line regions within AGN needs to be carried out, much like the work done for the broad line region and its filling factor \citep[e.g][]{Osterbrock1991}. 

\begin{table}
	\caption{Lower distance limits of the emission line regions in NGC 5548 and NGC 3783, using the results from \cite{Mao5548} and \cite{Mao3783}, respectively. The distances are for the narrow line components (N1 and N2), and the broad components (B1). See Section \ref{AGN_Comp} for details.}
	\label{Dist_5547_3783}
	\centering
	\begin{tabular}{c | c c c}
	\hline
	\hline
	\multicolumn{0}{c}{} & N1 (pc) & N2 (pc) & B1 (pc) \Tstrut\Bstrut\\
	\hline
	Model\ \ \tablefootmark{a} & \multicolumn{3}{c}{NGC 5548} \Tstrut\Bstrut\\
	\hline
	D & 13 & 142 & - \Tstrut\Bstrut\\
	T & 19 & 93 & 0.09 \Tstrut\Bstrut\\
	\hline
	Date\ \ \tablefootmark{b} &  \multicolumn{3}{c}{NGC 3783} \Tstrut\Bstrut\\
	\hline
	2000/01 & 0.01 & 1.94 & 0.01 \Tstrut\Bstrut\\
	11 Dec 2016 & 0.03 & 10.4 & 0.01 \Tstrut\Bstrut\\
	21 Dec 2016 & 0.04 & 15.6 & 0.01 \Tstrut\Bstrut\\
	 \hline              
	\end{tabular}\\
\tablefoot{
\tablefoottext{a}{The two models used on NGC 5548 by \cite{Mao5548} from the 2013/14 observations: \textit{D} for two emission (narrow) components; \textit{T} for three emission componenets.}
\tablefoottext{b}{The observation dates when the data of NGC 3783 were obtained and used in \cite{Mao3783}.}
}
\end{table}

\section{Conclusion}
\label{Conclusion}
We have investigated the high resolution, RGS spectrum of NGC 7469, using the photoionisation model \texttt{PION}, in \texttt{SPEX}, to characterise both the emission line regions and the WAs. For the first time in NGC 7469, limits on the distances of the narrow emission line regions from the central black hole have been estimated. The main conclusions from this investigation are detailed as follows:

\begin{itemize}

	\item{The emission line regions are found to be made up of three components: two narrow line and one broad line components. The two narrow emission components (log $\xi =$ 0.4 and 1.6; log $N_H =$ 27.8 and 26.7, respectively) were found to fit the majority of the emission features.}
	
	\item{Assuming a volume filling factor value of 0.1, the two narrow components, EM1 and EM2, are found to be located at 2.6 and 2.5 pc away from the central source, respectively.}
	
	\item{A broad emission component, with an outflow velocity of $v_{out} = -4460$ km s\textsuperscript{-1} and broadening velocity of $v_b = 1360$ km s\textsuperscript{-1} \citep[similar to the optical BLR values;][]{Suganuma2006, Kollatschny2013}, is found to reproduce emission at the position of the intercombination line of the \ion{O}{VII} triplet. However, we found that this broad component is actually the forbidden line, blueshifted to higher energies, coincidentally filling in the intercombination line. The distance of EM3 is found at either 0.03 pc (assuming a volume filling factor of 0.001) or 0.004 pc, depending if the ionisation parameter or the outflow velocity is used in the distance determination, respectively.}
	
	\item{The RGS spectrum also shows emission features from collisionally ionised plasma, produced in the starburst region of NGC 7469. Here, the electron temperature is found to be 0.32 keV, with the plasma moving towards us at -282 km s\textsuperscript{-1}, consistent with \cite{Behar2017, Mehdipour2018}. }
	
	\item{We find the WA is explained by three ionisation (log $\xi =$ 2.3, 3.0, and 1.6) and three kinematic ($v_{out} = $ -630, -910, and -1960 km s\textsuperscript{-1}) phases, with the highest outflow velocity having the smallest $\xi$ \citep{Blustin2003, Kriss2003}. The WA kinematics are also similar to the UV velocity components from this campaign \citep{Aravprep} and from archive results \citep{Kriss2003, Scott2005}.}
	
	\item{We find that the total column density of the WA is $N_H^{Tot} = 64 \times 10^{24}$ m\textsuperscript{-1}, about twice as large as the total found by \cite{Behar2017}.}
	
	\item{The upper distances of the WA components are found to be 1.9, 0.3, and 0.6 pc for components 1 to 3, respectively, consistent with previous findings \citep{Blustin2007}. The location of the torus is estimated at around 1.2 pc.}
	
	\item {We find that it is very unlikely that any of the plasma components are made up of both emission and absorption regions. This is due to the large uncertainties associated with the respective volume filling factors and different distance measurements of the WA components from this work and from variability arguments \citep[e.g.][]{Mehdipour2018}.}
	
	\item{The total mass outflow rate of all three WA components combined is 0.042 $M_{\sun}$ yr\textsuperscript{-1}, and the kinematic luminosities are measured between $\log L_K = 32$ and $\log L_K = 33$ W. The fraction of the kinematic luminosity compared to the bolometric luminosity \citep[$L_{bol} = 2.5 \times 10^{37}$ W;][]{Petrucci2004} is $\sim 1 - 3 \times 10^{-3} \%$, similar to the fractions found by \cite{Blustin2007}.}
	
	\item{To compare the emission line regions of NGC 7469 with those in other AGN, we turned to NGC 5548 \citep{Mao5548} and NGC 3783 \citep{Mao3783}, which were both analysed with the photoionisation model \texttt{PION}. For NGC 5548, we calculate distances which are comparable to previous analysis of the NLR \citep{Whewell2015} and we find that the overall structure within the AGN of NGC 5548 is similar to that of NGC 7469. For NGC 3783 on the other hand, the first narrow emission line component, due to its very large ionisation parameter, is found to be at a similar distance to the broad emission component. This is at variance with the expectations of standard AGN models where the NLRs are further out from the nucleus than the BLR.}

	\item{This problem is overcome by allowing the volume filling factor to be between 0.1 and 1 for the emission line regions. This allows the NLR distances in NGC 7469 and NGC 5548 \citep{Whewell2015} to be further out than the WA, and would also imply that the first narrow emission component in NGC 3783 could be 10 times further away from the central source than the broad component. Alternatively, the volume filling factor of the broad component in NGC 3783 may be $C_v \sim 0.0001$, which would allow the first narrow emission component to be further away than the broad component. Despite the fact that this may seem an ad hoc solution, further detailed investigations of the volume filling factor in narrow emission line regions are necessary in order to provide firmer ground to distance estimations in AGN.}
\end{itemize}

	\begin{acknowledgements}
	This work is based on observations obtained with XMM-Newton, an ESA science mission with instruments and contributions directly funded by ESA Member States and NASA. SGW acknowledges the support of a PhD studentship awarded by the UK Science \& Technology Facilities Council (STFC). SB acknowledges financial support from the Italian Space Agency under grant ASI-INAF 2017-14-H.O. BDM acknowledges support from the European Union’s Horizon 2020 research and innovation programme under the Marie Skłodowska-Curie grant agreement No 798726. JM is supported by the University of Strathclyde UK APAP network grant ST/R000743/1. POP thanks the French CNES agency for financial support.
\end{acknowledgements}


	\newpage
	\begin{appendix}
		\section{The velocity structure of the Galaxy}
		\label{Appendix}
		
		We modelled the absorbing material of the Galaxy using \texttt{HOT} in \texttt{SPEX} (see Section \ref{Galactic_Absorption}), with the sum of \ion{H}{I} and \ion{H}{II} column densities initially set to $5.5 \times 10^{24}$ m\textsuperscript{-2} \citep{Wakker2006, Wakker2011}, and a temperature of 0.5 eV to mimic neutral gas. However, the velocity structure of the gas in the Galaxy was unknown, meaning we did not have any information about the turbulent velocity for the \texttt{HOT} components. 
		
\begin{figure}
			\centering
			\includegraphics[width=\linewidth]{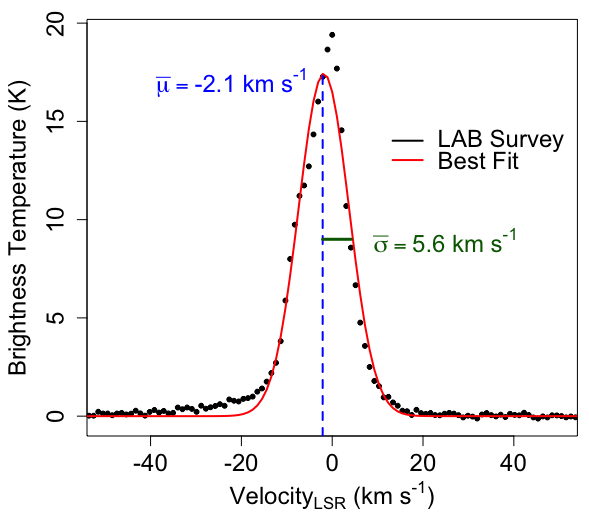}
			\caption{Galactic 21 cm \ion{H}{I} profile in our line of sight towards NGC 7469, of brightness temperature as a function of velocity, from the \texttt{LAB} survey (black dots). The red line is the Gaussian function (Equation \ref{Gaussian}) used to determine the velocity dispersion $\sigma_v$ and the velocity mean $\mu$. Estimates of these two parameters were obtained by averaging the values from the \texttt{LAB} \citep{LAB2005} and \texttt{EBHIS} \citep{EBHIS2016} surveys. The average velocity dispersion is used as the turbulent velocity for the Galaxy in the \texttt{HOT} components; see text for details of this argument.}
			\label{Galaxy_Profile}
		\end{figure}

		To determine the turbulence of the neutral (and mildly ionised) Galaxy absorbing gas in the line of sight towards NGC 7469, we used the 21 cm \ion{H}{I} profiles from the \texttt{EBHIS} \citep{EBHIS2016} and \texttt{LAB} \citep{LAB2005} surveys\footnote{The interface to measure the \ion{H}{I} profiles can be found at \url{https://www.astro.uni-bonn.de/hisurvey/AllSky_profiles/index.php}}. Figure \ref{Galaxy_Profile} shows the data from the \texttt{LAB} survey (black dots), displaying the 21 cm \ion{H}{I} profile in terms of brightness temperature ($T_B$) as a function of velocity at the local standard of rest ($V_{LSR}$).
		
		In order to obtain the information regarding the velocity, we fitted a Gaussian to the survey data, red line in Figure \ref{Galaxy_Profile}, given by
		\begin{equation}
		T_B = A \exp{\left(-\frac{(v - \mu)^2}{2 \sigma_v^2}\right)},
		\label{Gaussian}
		\end{equation}
		where $v$ is the velocity ($V_{LSR}$ data from the two surveys) and the standard deviation ($\sigma_v$), mean ($\mu$), and normalisation (A) are the three unknown parameters. We did this for both \texttt{LAB} and \texttt{EBHIS} surveys and averaged the three unknown parameters $\bar{\sigma}$, $\bar{\mu}$ and $\bar{A}$.  The average mean and normalisation values were $\bar{\mu} = -2.1 \pm 0.9$ km s\textsuperscript{-1} and $\bar{A} = 16.9 \pm 2.3$ K respectively. The average velocity dispersion is found to be $\bar{\sigma} = 5.64 \pm 1.10$ km s\textsuperscript{-1}.
		The \ion{H}{I} column density ($N_H$) is found using the \texttt{LAB} survey \citep{Wakker2011}, which means the $\bar{\sigma}$ value is needed from the \ion{H}{I} profile of $N_H(v)$ to get the turbulent velocity in the Galaxy. However, the \ion{H}{I} profile for $N_H(v)$ is, in fact, the same as the profile for $T_B(v)$, as the following argument proves \citep{Chengalur2013}:
		\begin{equation}
		T_B(v) = T_S(v)[1-e^{-\tau(v)}]
		\label{EQ_TB}
		\end{equation}
		\begin{equation}
		N_H(v) = K \int{T_s(v) \tau(v)} dv
		\label{EQ_NH}
		\end{equation}
		In these two equations, $T_B(v)$ is the brightness temperature (Figue \ref{Galaxy_Profile}), $T_S(v)$ is the hydrogen spin temperature, $\tau(v)$ is the optical depth, and $K = 1.8 \times 10^{18}$ is a constant. Here, it is clear $N_H(v) \propto \tau(v)$, meaning the optical depth, measured via absorption measurements, is used to determine the column density \citep{Lee2015}. Substituting Equation \ref{EQ_TB} into Equation \ref{EQ_NH} to remove $T_S$ gives
		\begin{equation}
		N_H(v) = K \int{\frac{T_B(v) \tau(v)}{1-e^{-\tau(v)}}dv}.
		\end{equation}
		To estimate $N_H(v)$, we can assume that the gas in the Galaxy is optically thin ($\tau(v) << 1$) \citep{Lee2015}, which gives the final equation
		\begin{equation}
		N_H(v) = K \int{T_B(v)} dv,
		\label{EQ_NH_TB}
		\end{equation}
		meaning that the column density can be measured by integrating the emission spectrum of the 21 cm line ($T_B(v)$) over the velocity interval $dv$ ($V_{LSR}$). Equation \ref{EQ_NH_TB} implies $N_H(v) \propto T_B(v)$, so the \ion{H}{I} profile in Figure \ref{Galaxy_Profile} is the same as the \ion{H}{I} profile for $N_H(v)$, except with the factor of K being the difference \citep{Chengalur2013, Lee2015}. Therefore, the $\bar{\sigma}$ found from Figure \ref{Galaxy_Profile} can be taken as the velocity dispersion of the \ion{H}{I} profile of $N_H(v)$, which we use in the \texttt{HOT} models as the turbulent velocity of the Galaxy.
	\end{appendix}

\end{document}